%% file: EffPotHbExpPDF.tex
\pdfoutput=1
\pdfmapfile{+XCharter.map}
\documentclass[10pt,a4,twocolumn,twoside]{article}

\usepackage{./styling/style-main}

\title{On the relationship between gauge dependence and IR divergences in the $ \hbar $-expansion of the effective potential}
\author{Andreas~Ekstedt\thanks{andreas.ekstedt@physics.uu.se}\,}
\author{Johan~L\"ofgren\thanks{johan.lofgren@physics.uu.se}}
\affil{Department of Physics and Astronomy, Uppsala~University\\Box~516, SE-751~20~Uppsala, Sweden}
\date{\today}

\begin{document}
	\maketitle
	\thispagestyle{plain}

	\begin{abstract}
			Perturbative calculations of the effective potential evaluated at a broken minimum, $ V_{\text{min}} $, are plagued by difficulties. It is hard to get a finite and gauge invariant result for $ V_{\text{min}} $. In fact, the methods proposed to deal with gauge dependence and IR divergences are orthogonal in their approaches. Gauge dependence is dealt with through the $ \hbar $-expansion, which establishes and maintains a strict loop-order separation of terms. On the other hand, IR divergences seem to require a resummation that mixes the different loop orders. In this paper we test these methods on Fermi gauge Abelian Higgs at two loops. We find that the resummation procedure is not capable of removing all divergences. Surprisingly, the $ \hbar $-expansion seems to be able to deal with both the divergences and the gauge dependence. In order to isolate the physical part of $ V_{\text{min}} $, we are guided by the separation of scales that motivated the resummation procedure; the key result is that only hard momentum modes contribute to  $ V_{\text{min}} $.
	\end{abstract}
	\input{./tex/Introduction}
	\input{./tex/background}
	\input{./tex/results}
	\input{./tex/discussion}

	\input{./tex/acknowledgments}

	\appendix
	\input{./tex/conventions}
	\input{./tex/ward}
	\input{./tex/2LoopPot}
	\input{./tex/LDS}

	\let\enquote\letmacundefined 
	\bibliographystyle{bibstyle}
	{\bibFont
	\bibliography{EffPotHbExpPDF}%
	}%
\end{document}

%% file: tex/Introduction.tex
\section{Introduction}
One of the most important tools for studying spontaneous symmetry breaking within \qft{} is the effective potential $ V $, which can be considered as the quantum-corrected version of the classical potential $ V_{0} $. The effective potential is given by $ V=V_{0}+\hbar V_{1}+\hbar^{2} V_{2}+\mathellipsis $
with factors of $ \hbar $ inserted in order to emphasize that this quantity is usually calculated perturbatively. If the theory allows for spontaneous symmetry breaking through the scalar field $ \phi $, its vacuum expectation value, or \vev{}, would be found by extremizing the effective potential $ \partial V|_{\pb=\pbmin}=0$. In this way the effective potential allows us to find the quantum corrected minimum, $\pbmin$, and the corresponding background energy density $\Vmin \define V(\pbmin) $.

It can be hard to extract physical information from the effective potential. In particular, $ \Vmin $ is in principle a measurable quantity, yet there are difficulties in obtaining a physical value of $ \Vmin $ in perturbation theory.

One difficulty is gauge dependence. The effective potential is in general gauge dependent, but is guaranteed to be gauge independent when evaluated at its extremum $\pbmin$. However, as recently pointed out by Andreassen, Frost, and Schwartz\room\cite{Andreassen:2014eha}, and by Patel and Ramsey-Musolf\room\cite{Patel:2011th}, gauge invariance of $ \Vmin $ relies on a strict $ \hbar $ power counting. To establish a strict counting the aptly named $ \hbar $-expansion is used.

Another difficulty is that, at higher loop orders, the effective potential diverges near the broken minimum. These divergences come from Goldstone bosons becoming massless and signal a breakdown of the perturbative expansion. It has been proposed that a resummation might be required to obtain a finite result. Resummation of \ir{} divergences has been discussed by Martin in\room\cite{Martin:2014bca}, and by Elias-Mir\'o, Espinosa, and Konstandin in\room\cite{Elias-Miro:2014pca}\nocite{Coleman}. An extension of these methods to general Fermi gauges has been discussed in\room\cite{Espinosa:2016uaw}.

Both the gauge and \ir{} problems relate to the perturbative expansion, but the solution of the gauge invariance and \ir{} divergence issues stand in contrast to each other. Gauge invariance requires a strict separation of loop orders, and \ir{} divergences suggest that contributions from all loop orders should be included.
 Additionally, the resummation procedure of\room\cite{Espinosa:2016uaw} is problematic in the presence of certain gauge dependent singularities. These "new" singularities can naviely not be resummed. We argue that in a general model with spontaneous symmetry breaking, the $ \hbar $-expansion is capable of treating both the gauge invariance and the \ir{} divergence issues. We demonstrate this in section\room\ref{sec:results}, with the help of the momentum separation techniques of\room\cite{Elias-Miro:2014pca,Espinosa:2016uaw,Espinosa:2017aew}.

In section\room\ref{sec:background} relevant notation is introduced and the theoretical background is reviewed. Section\room\ref{sec:results} summarizes our main results. To illustrate our procedure, the full 2-loop effective potential is calculated in the Abelian Higgs model. Details and proofs can be found in the appendix. Finally, conclusions are given in section\room\ref{sec:disc}.

%% file: tex/background.tex
\section{Background}\label{sec:background}
The section starts with a summary of conventions, presented for the Abelian Higgs model. We review the origins of gauge dependence and \ir{} divergences. Methods for dealing with these issues are discussed: a consistent $\hbar$-expansion and daisy resummation respectively. The general conventions are then collected in appendix\room\ref{app:conventions}.
\subsection{Abelian Higgs}\label{ssec:AH}
The Abelian Higgs model is a useful toy-model because it exhibits the issues that we want to discuss: gauge dependence and \ir{} divergences. The model consists of a \uone{} gauge field $ A^{\mu} $ together with a complex scalar $ \Phi =\frac{1}{\sqrt{2}}\left(\phi_{1}+i\phi_2\right)$ charged under this symmetry. The Lagrangian, using the conventions of\room\cite{Andreassen:2014eha}, is
\begin{align*}
\begin{split}
	\lag=&\lag_{\textcaps{AH}}+\lag_{\mathrm{g.f.}}+\lag_{\mathrm{ghost}},
	 \\\lag_{\textcaps{AH}}=&-\frac{1}{4}F_{\mu \nu}F^{\mu \nu}-\left(D^{\mu}\Phi\right)^{\dagger}D_{\mu}\Phi - V_{0}\left[\Phi,\Phi^{\dagger}\right]\\
\end{split}
\end{align*}
where $ F_{\mu \nu}=\partial_{\mu}A_{\nu}-\partial_{\nu}A_{\mu} $ is the \uone{} field strength, $ D_{\mu}=\partial_{\mu}+i e A_{\mu} $ is the covariant derivative, $ \lag_{\mathrm{g.f.}}+\lag_{\mathrm{ghost}}$ contains the details of gauge fixing and the ghost sector, and $ V_{0}[\Phi,\Phi^{\dagger}] $ is the classical scalar potential. Expressed in terms of the real degrees of freedom $\smash{ \vec{\phi}=\left(\phi_{1},\phi_{2}\right)} $, the classical potential, $ V_{0}$, is
\begin{equation*}
	V_{0}\left[\phi_{1},\phi_2\right]=-\frac{1}{2}m^{2}\left(\phi_{1}^2+\phi_2^2\right)+\frac{1}{4!}\lambda\left(\phi_{1}^2+\phi_2^2\right)^{2}.
\end{equation*}

The Lagrangian, $\lag_{\textcaps{AH}}$, is invariant under global and local \uone{} transformations. However, the remaining terms $ \lag_{\mathrm{g.f.}}, \lag_{\mathrm{ghost}} $, explicitly break the gauge invariance. Common gauge fixing choices are discussed in\room\cite{Andreassen:2014eha}; we note that the commonly used $ R_{\xi} $ gauges,
\begin{align*}
	\lag_{\mathrm{g.f.}}&=-\frac{1}{2 \xi} \left(\partial_{\mu} A^{\mu}+\xi \pb \phi_{2}\right)^{2},\\
	\lag_{\mathrm{ghost}}&=-\bar{c}\left(\partial^2-\xi e^2 \phi^2\left(1+\frac{h}{\phi}\right)\right)c,
\end{align*}
explicitly break the remnant global \uone{} symmetry. The breaking of this symmetry complicates calculations involving Goldstone bosons. In this paper we are interested in the interplay between \ir{} divergences and gauge dependence of the effective potential. These features are most evident in Fermi gauges, which will be used in the remainder of this paper. In these gauges, the gauge fixing and ghost terms are
\begin{align*}
	\lag_{\mathrm{g.f.}}&=-\frac{1}{2 \xi} \left(\partial_{\mu} A^{\mu}\right)^{2},\\
	\lag_{\mathrm{ghost}}&=-\overline{c}\partial^{\mu}\partial_{\mu} c.
\end{align*}
In Fermi gauges ghosts are free. A slight complication with these gauges is kinetic mixing between the longitudinal gauge boson mode and the Goldstone boson.

Because the potential is invariant under the global \uone{} symmetry, we have the freedom to place the \vev{} of our vector $ \vec{\phi} $ in any direction. Choosing the \vev{} $ \pb $ to lie in the $ \phi_{1} $ direction, that is, shifting $ \phi_{1}\rightarrow\pb+\phi_{1} $, the field-dependent masses squared in the presence of the background field are
\begin{align*}
	H&=-m^{2}+\frac{1}{2}\lambda \pb^{2},\\
	G&=-m^{2}+\frac{1}{6}\lambda \pb^{2},\\
	A&=e^{2}\pb^{2}.
\end{align*}
The masses denote the tree-level mass of the fields: the Higgs mass $ H $, the Goldstone mass $ G $, and the "photon" mass $A$ --- with the notation that the mass-squared of field $ X $ is also denoted as $ X $. Due to the kinetic mixing between the longitudinal mode of $ A^{\mu} $ and $ G $, it is useful to introduce the masses $ G_{+} $ and $ G_{-} $,
\begin{equation*}
	G_{\pm}=\frac{1}{2}\left(G\pm\sqrt{G\left(G-4\xi A\right)}\right).
\end{equation*}
 The masses $ G_{+} $ and $ G_{-} $ depend explicitly on the gauge fixing parameter $ \xi $.

There are different possible scenarios depending on the value of $ m^{2} $; it is assumed that $ \lambda>0 $.
\begin{itemize}
	\item $ m^{2}<0 $: This model is called scalar \qed{}. Scalar \qed{} consist of two scalars, with squared mass $ \left|m^{2}\right| $, and a massless gauge boson. This model is not considered in this work because there is no spontaneous symmetry breaking.
	\item $ m^{2}=0$: This is the Coleman-Weinberg (\cw{}) model. Classically, this model does not exhibit spontaneous symmetry breaking, but masses can be generated through quantum corrections. The generation of mass needs a careful treatment of perturbation theory --- the coupling $ \lambda $ neccesarily scales as $ \hbar $, as is discussed in\room\cite{Andreassen:2014eha}. The model is discussed in subsubsection\room\ref{sssec:cw}.
	\item $\smash{ m^{2}>0} $: This model is called Abelian Higgs. There is spontaneous symmetry breaking because the classical potential has a minimum located at $\smash{\pb_{0}=\sqrt{6 m^{2}/\lambda }}$; the masses evaluated at this field point are
	\begin{align*}
		\left.H\right|_{\pb_{0}}&=2m^{2},\\
		\left.G_{\pm}\right|_{\pb_{0}}&=0,\\
		\left.A\right|_{\pb_{0}}&=6 \frac{e^{2}}{\lambda}m^{2}.
	\end{align*}
\end{itemize}
\vspace*{-1\leadingHeight}
The Abelian Higgs model is the main focus in this paper. The Feynman rules relevant for deriving the effective potential are given in appendix\room\ref{app:conventions}.
\subsection{The effective potential}\label{sec:effpot}
In this subsection conventions for the 1-loop potential are given. The $ \hbar $-expansion and daisy resummation methods are reviewed.
\subsubsection{The 1-loop potential}
The 1-loop contribution of a scalar field with mass $X$ is in \msbar{} given by\footnote{For a more complete discussion of the 1-loop potential we recommend\room\cite{Andreassen:2014eha}.}
\begin{align}\label{eq:v1}
 V_{1}(\pb) \sim \frac{1}{4} X^{2}\left(L_{X}-\frac{3}{2}\right),
\end{align}
where we introduced the same shorthand as in\room\cite{Espinosa:2016uaw}, $ \LOG{X}\define \log\left[X / \mu^{2}\right] $ with $ \mu $ the \msbar{} renormalization scale. To simplify the formula, $ \hbar $ is rescaled with a factor of $ 16 \pi^{2} $.
\subsubsection{The $ \hbar $-expansion}
 Though the Nielsen identity\room\cite{Nielsen:1975fs} guarantees that the physical quantity $ \Vmin $ is gauge invariant, care must be taken when $ \Vmin $ is calculated in perturbation theory. The issue is that the Nielsen identity is a non-perturbative statement, but in perturbation theory things are more subtle.

A consistent $ \hbar $-expansion is necessary in order to establish this gauge invariance. This $ \hbar $-expansion has recently been discussed by Patel and Ramsey-Musolf in\room\cite{Patel:2011th}, but see also \room\cite{Laine:1994zq} and\room\cite{Farakos:1994xh} for earlier applications. The key point is in how the minimum $ \pbmin $ is treated. The minimum is found by solving the equation
\begin{equation}\label{eq:diffV}
\left.\partial V\right|_{\pb=\pbmin}=0.
\end{equation}
Because the potential $ V $ is calculated perturbatively, equation\room\eqref{eq:diffV} should also be solved perturbatively, order by order in $ \hbar $. This gives $ \pbmin=\pbmin_{0}+\hbar \pbmin_{1}+\hbar^{2}\pbmin_{2}+\mathellipsis $, where the contributions $ \pbmin_{1},\pbmin_{2},\mathellipsis $, are found by inserting this expansion in equation\room\eqref{eq:diffV},
\begin{align*}
\left.\partial V_{0}\right|_{\pbmin_{0}}&=0,\\
\left.\partial V_{1}\right|_{\pbmin_{0}}+	\pbmin_{1}\left.\partial^{2} V_{0}\right|_{\pbmin_{0}}&=0\implies \pbmin_{1}=-\left.\frac{\partial V_{1}}{\partial^{2} V_{0}}\right|_{\pbmin_{0}}.\\
\vdots
\end{align*}
When $ \pbmin $ is evaluated perturbatively, $ \Vmin $ can be consistently calculated order by order in perturbation theory. The first few terms of $ \Vmin $ are
\begingroup
\small
\begin{equation*}
V(\pbmin)=\left.V_{0}\right|_{\pbmin_{0}}+\hbar \left.V_{1}\right|_{\pbmin_{0}}+\hbar^{2}\left.\left(V_{2}-\frac{1}{2}\left(\pbmin_{1}\right)^{2}\partial^{2}V_{0}\right)\right|_{\pbmin_{0}}+\mathellipsis
\end{equation*}
\endgroup
It has been shown that $V(\pbmin)$ evaluated in this way is gauge invariant order by order in $ \hbar $\room\cite{Patel:2011th}. In the Fermi gauges it can be shown that $ \pbmin_{1} $ and higher order contributions have \ir{} divergences of the form $ \xi \log{\left[G\right]} $ and worse\room\cite{Espinosa:2016uaw}. However, these divergences are not a problem because the \vev{}
$\pbmin{}$ is not a physical quantity. Furthermore, the divergence of $ \pbmin $ ensures that the effective potential is finite. All gauge dependent divergences are guaranteed to cancel order by order in $ \hbar $.

As an aside we would like to comment on the term "$\hbar$-expansion", of which there are contradictory uses. It has historically been thought that when the units are changed from natural units, and explicit factors of $\hbar$ are reinserted, that these factors of $\hbar$ act as loop counting parameters. This is not true. As was emphasized in \cite{Holstein:2004dn} it is possible to have loop effects at the order $\hbar^0$. That is, it is possible to calculate classical corrections by doing loops. The main message is that this kind of $\hbar$-expansion is not a loop-expansion. In this paper we do not reintroduce factors of $\hbar$ by changing units. Instead, we use $\hbar$ as the name of the loop-counting parameter---the perturbative expansion is then called the "$\hbar$-expansion". As far as we can tell, this nomenclature was introduced in\room\cite{Patel:2011th}. It is unfortunate that both of these kinds of expansions are known as $\hbar$-expansions.

In order to clarify the nature of the perturbative expansion we are using, we ask the reader to consider the one-loop correction to the Higgs mass
\begin{align*}
m_\text{H}^2=m_{\text{tree}}^2+\hbar~\Pi_1(m_\text{H}^2),
\end{align*}
where $m_{\text{tree}}$ is the tree level Higgs mass and $\Pi_1(p^2)$ is the one-loop self energy. To find the one-loop correction to the Higgs mass we should expand $m_\text{H}^2=m_{\text{tree}}^2+\hbar~m_1^2+\ldots$, and would find $m_1^2=\Pi_1(m_{\text{tree}}^2)$. This is the same perturbative matching that we will use throughout this paper when discussing the effective potential.

\subsubsection{IR divergences in the effective potential}\label{ssec:IR}
The effective potential evaluated at $ \pb_{0} $ has \ir{} divergences when calculated perturbatively (because the Goldstone bosons are massless). These \ir{} divergences point to a problem with the perturbative expansion, suggesting that it might be necessary to perform a resummation. To illustrate the idea, consider Landau gauge --- where the mixed $ G $-$ A^{\mu} $ propagator vanishes, and the Goldstone propagator is $ \smash{D_{G}(k)=i/(k^{2}-G) }$. The most severe divergences are given by daisy diagrams\room\cite{Martin:2014bca,Elias-Miro:2014pca},
\begin{equation*}
	\frac{1}{2}
	\begin{gathered}
		\includegraphics[height=30pt]{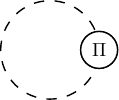}
	\end{gathered}\room,
	\frac{1}{4}
	\begin{gathered}
		\includegraphics[height=30pt]{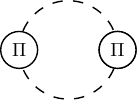}
	\end{gathered}\room,
	\frac{1}{6}
	\begin{gathered}
		\includegraphics[height=30pt]{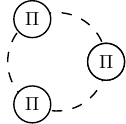}
	\end{gathered}\room,
	\begin{gathered}
		\mathellipsis
	\end{gathered}
\end{equation*}
The daisies are built from Goldstone propagators and insertions of the Goldstone self-energy $ \Pi(k^{2}) $. These diagrams, for a given loop order L, contain the worst divergences in the limit $ G\rightarrow 0 $ because they have the maximum number of Goldstone propagators within the same loop. The value of a daisy diagram with $ n $ "petals" can be written
\begin{equation*}
\-\frac{1}{2 n}\momint{k} \left(\frac{1}{k^{2}-G}\right)^{n}(-\Pi(k^{2}))^{n},
\end{equation*}
where we introduced a shorthand notation for the integration, $ \momint{q}=Q^{2\epsilon}\int \dif^{d}q/\left(2 \pi\right)^{d} $, with $ Q $ the renormalization scale, and $ d=4-2\epsilon $. \ir{} divergences come from the limit $G\rightarrow 0$. The daisy integrand, for \emph{soft} momenta $k^{2} \sim G $, scales as $ \sim G^2\left(\Pi(0)/G\right)^{n} $. The daisy diagrams are \ir{} divergent for $ n \geq 2$, which corresponds to 3 loops and higher. In this argument we used $ \Pi(0) $ instead of $ \Pi(k^2) $, because for soft momenta the momentum dependence of $ \Pi(k^2) $ simply corresponds to sub-divergences,
\begin{align*}
\Pi(k^2)\sim \Pi(0)+G \Pi^{'}(0)+\mathellipsis
\end{align*}

Because \ir{} divergences come from momenta of the order $k^2\sim G$ it's useful to separate different contributions as hard and soft\room\cite{Elias-Miro:2014pca,Espinosa:2017aew}. Fields with momentum that scale as $ k^{2}\sim G $ are soft; fields with momentum that scale as $ k^{2} \gg G $ are hard. It is then possible to separate the effective potential into a hard and a soft part, $V(\pb)=\VHard{}(\pb)+\VSoft{}(\pb)$, where $ \VHard $ only contains contributions from hard fields, and $ \VSoft $ from soft and hard. We refer to this separation as the hard/soft split. From $	\overline{G}=\frac{1}{\phi}\partial V(\phi)$ the Goldstone self-energy can be separated into a hard part $ \Delta $ and a soft part $ \Sigma $, $ \Pi(0) = \Delta + \Sigma $, with $ \Delta = \frac{1}{\phi}\partial \VHard $, $ \Sigma = \frac{1}{\phi}\partial \VSoft $.

At one loop the hard/soft splitting is simple. For Abelian Higgs in Landau gauge the potential splits as
\begin{align*}
\VHard_{1}(\pb)&=\frac{1}{4}\left[H^{2}\left(\LOG{H}-\frac{3}{2}\right)+3 A^{2}\left(\LOG{A}-\frac{5}{6}\right)\right],\\
\VSoft_{1}(\pb)&=\frac{1}{4} G^{2}\left(\LOG{G}-\frac{3}{2}\right).
\end{align*}

The Goldstone self-energy splits as
\begin{align}
\Delta_{1}&=\frac{1}{2}\left[\lambda H\left(\LOG{H}-1\right)+3 e^{2}A\left(\LOG{A}-\frac{1}{3}\right)\right],\label{eq:Delta1AH}\\
\Sigma_{1}&=\frac{1}{6}\lambda G\left(\LOG{G}-1\right).
\end{align}
Because \ir{} divergences come from soft momenta, all \ir{} divergences of the effective potential ($ \overline{G} $) are contained in $ \VSoft$ ($ \phi  $).

\subsubsection{Resummation in Landau gauge}
The idea of resummation is that when the Goldstone mass becomes small, corrections from all orders are relevant. The result of the leading daisy resummation is
\vspace*{-.25em}
\par\nobreak
\begingroup \small
\begin{align*}
\sum_{n=1}^\infty &\frac{1}{2 n}\momint{k}(-1)^{n+1} \left(\frac{1}{k^{2}-G}\right)^{n}(\Pi(0))^{n}=\\
&\frac{1}{2} \momint{k}\log{\left[k^{2}-G-\Pi(0)\right]}-\frac{1}{2}\momint{k}\log{\left[k^{2}-G\right]}.
\end{align*}
\endgroup
The Goldstone 1-loop contribution is $\frac{1}{2} \momint{k}\log\left[k^{2}-G\right] $: the resummation implies that the Goldstone mass in the 1-loop potential should be shifted according to $ G\rightarrow \overline{G}=G+\Pi(0)$.

Because $ \overline{G} $ is the inverse of the Goldstone two-point function evaluated at zero momentum, it is possible to calculate it from the effective potential,
\begin{equation}\label{eq:Gbar}
\overline{G}=\partial_{\phi_{2}}^{2}V(\phi_{1},\phi_{2})=\frac{1}{\phi}\partial V(\phi),
\end{equation}
where the last equality follows from the Goldstone theorem.
A framework for resumming \ir{} divergences was developed for Landau gauge in\room\cite{Elias-Miro:2014pca,Espinosa:2017aew}.

This framework relies on that only the hard part of the self-energy should be used in the resummation, that is
\begin{equation}\label{eq:GbarDelta}
\Gbar = G + \Delta.
\end{equation}
It was shown in\room\cite{Espinosa:2017aew} that this resummation is, in Landau gauge, consistent to all orders in perturbation theory. The argument of\room\cite{Espinosa:2017aew} relies on the fact that diagrams in which all particles have soft momenta scale as $\sim G^2$. Shifting $G\rightarrow \Gbar$ in these diagrams gives a finite result in the limits $ G \rightarrow 0,~ \Gbar \rightarrow 0 $.

The above described resummation works to all orders in Landau gauge. An attempt to extend the resummation method to also cover general Fermi gauges has been done in\room\cite{Espinosa:2016uaw}, but we find that there are certain complications. First, the hard self energy is in general gauge dependent at two-loops and higher. Second, the purely soft potential has \ir{} divergences. We demonstrate these complications in section\room\ref{sec:results}.

%% file: tex/results.tex
\section{Results}\label{sec:results}
We show that there are two different classes of \ir{} divergences in the Fermi gauges, and that the resummation procedure defined in equation\room\eqref{eq:GbarDelta} is incapable of dealing with all \ir{} divergences. We argue that the $\hbar$-expansion is free of \ir{} divergences to all orders.

\subsection{IR divergences in the fermi gauges}\label{ssec:powercountingfermi}
Let us discuss the origin of \ir{} divergences in Fermi gauges, to parallel how daisy diagrams are resummed in Landau gauge (see subsubsection\room\ref{ssec:IR}). Before turning to the Fermi gauges, it is useful to review how divergences appear in Landau gauge.
\subsubsection{Power counting in Landau gauge}
Our main concern are diagrams where the momenta on all lines are soft. These diagrams remain after the leading divergent diagrams with hard self-energies have been subtracted; it's important that these all-line-soft diagrams are finite in the limit $ \smash{G \rightarrow 0} $. The power of $ G $ that a generic soft vacuum diagram $ D $ scales with is denoted by $ P_{G}(D)|_{\xi=0} $. Denoting the number of $ A$-$G$-$H $ vertices $ V^{AGH} $, and similarly for other vertices, it can be shown that
\begin{align*}
\left.P_G(D)\right|_{\xi=0}=2&+V^{A G H}+V^{GG AA}+2 V^{HH AA}\\
&+V^{HH GG}+V^{HHH}+2 V^{HHHH}.
\end{align*}
$ \smash{P_{G}(D)|_{\xi=0}} $ shows that all-line-soft diagrams, at all loop levels, scale with a positive power of $ G $. It's then safe to shift $ G \rightarrow \Gbar $ in these diagrams.
\subsubsection{Power counting in the Fermi gauges}
The power counting is a bit more involved in the Fermi gauges. Consider a daisy diagram at $ L $ loops,
\vspace*{-.25em}
\par\nobreak
\begingroup \small
\begin{align}
\frac{1}{2}\frac{(-1)^{L}}{L-1}\int (d k) \left(\frac{k^2-\xi m_A^2}{(k^2-G_{+})(k^2-G_{-})}\right)^{L-1} \times \Delta^{L-1},
\end{align}
\endgroup
where $ \Delta $ is the hard part of the Goldstone self-energy, as described in subsubsection\room\ref{ssec:IR}. Here \ir{} divergences come from the momentum region $k^2\sim G_\pm \sim 0$. For soft momenta $ k^{2} \sim G_{\pm} $ this diagram scales as $ \sim\ (G_\pm)^2(G_\pm-\xi A)^{L-1} \Delta^{L-1}/(G_\pm^2)^{L-1} $. Logarithmic divergences now appear already at two loops, and are gauge dependent. At three loops there are divergences of the forms $ \xi^2/G_\pm^4 $, $ \xi/G_\pm^2 $ and $ \log G_\pm $. The 3-loop \ir{} divergence proportional to $\log G_\pm$ is not proportional to $\xi$ and is the same divergence we found in Landau gauge.

Just as in Landau gauge, the powers of $G_\pm$ for a generic diagram can be found through power counting. Note that some propagators (photon, Goldstone, mixed) have two terms with different scaling, and that the most severely \ir{} divergent terms are those that scale with the lowest power of $ G_{\pm} $. We will for the moment focus on diagrams with the most severe \ir{} divergences. These diagrams scale with $ G_{\pm} $ as%
\vspace*{-.25em}
\par\nobreak
\begingroup \small
\begin{equation*}
P_{G_{\pm}}(D)=2-2V^{gggg}-2V^{ggh}-V^{gg\gamma\gamma}+V^{hhh}+2V^{hhhh}+V^{hh\gamma\gamma}.
\end{equation*}
\endgroup
Because some of the vertices contribute negatively to $ P_{G_{\pm}}(D) $, it's possible to have diagrams that scale with non-positive powers of $ G_{\pm} $. The powers of $G_\pm$ can be reshuffled to show the interplay between \ir{} divergences and gauge dependence,
\begin{align}
P_{G_{\pm}}(D)=P_{G}(D)|_{\xi=0}-N_\xi(D),\label{eq:xiPG}
\end{align}
where $N_\xi (D)$ denotes the power of $\xi$ in the most severely \ir{} divergent term in diagram $ D $. The class of diagrams discussed are possible at two loops or higher. Because they are not guaranteed to be finite in the $ G_{\pm} \rightarrow 0 $ limit, it's of no use to shift $ G_{\pm} $ with $\Gbar_{\pm} $. This implies that, in its current form, the resummation procedure is not able to remove all \ir{} divergences. In subsubsection\room\ref{sssec:AH} we show a specific diagram where this happens.

The gauge dependent \ir{} divergences can be thought of as artifacts from neglecting a proper perturbative expansion. From equation\room\eqref{eq:xiPG} we note that this type of divergences is necessarily gauge dependent. Hence, these divergences are guaranteed to cancel when $ \Vmin $  is evaluated order by order in $ \hbar $.

To summarize, there are two classes of \ir{} divergences in the effective potential. The \ir{} divergences in the first class are gauge dependent and cancel when $ \Vmin $ is evaluated order by order in $ \hbar$. This class of \ir{} divergences is not resummable. The second class of \ir{} divergences consists of independent\footnote{Gauge independent in the sense that they do not depend on $ \xi $.} \ir{} divergences which are present in Landau gauge, and can be resummed. \emph{A priori}, it is not clear how this second class fares under the $ \hbar $-expansion.

These two classes of \ir{} divergences seem to require different methods to deal with them. We explore this in the next subsection.
\subsection{A strategy for calculating $ V_{\textrm{min}} $}\label{ssec:strategy}
Because there are gauge dependent \ir{} divergences in Fermi gauges that can not be removed through resummation, the need for resumming the effective potential is not evident. However, such gauge dependent divergences are guaranteed to cancel when evaluating $ \Vmin $ perturbatively through the $ \hbar $-expansion. We expect that the remaining \ir{} divergences also cancel in the $\hbar$-expansion.

In this subsection, using the hard/soft split introduced in\room\cite{Elias-Miro:2014pca,Espinosa:2016uaw,Espinosa:2017aew}, we prove that the three loop evaluation of $ \Vmin $ through the $ \hbar $-expansion is finite. We also show that all the \emph{leading} singularities cancel to all orders. We argue that the $ \hbar $-expansion is \ir{} fit to all orders. As a result, all contributions coming from soft modes cancel: only hard modes are relevant when evaluating  $ \Vmin $.

\subsubsection{The hard/soft split and the $ \hbar $-expansion}
The separation of the effective potential into separate contributions from hard and soft momenta,
$ V = \VHard+\VSoft $, is valid for $ \pb \approx \pbmin_{0} $ such that there is a separation of scales between $ G $ and other heavy fields.

Because the goal is to calculate $ \Vmin=V(\pbmin) $, we separate $\Vmin$ into hard and soft contributions, $ \Vmin = \Vminh+\Vmins $\textemdash we emphazise that this split is only used as tool to track the \ir{} divergences. In order to isolate hard/soft contributions in an efficient way, we need to deal with the vev $ \pbmin=\pbmin_{0}+\hbar \pbmin_{1}+\mathellipsis $ in this separation. It is useful to split the vev as $ \pbmin = \pbh+\pbs $, where $ \pbh $ is defined through $ \left.\partial \VHard \right|_{\pb=\pbh}=0$, and $ \pbs $ denotes a remainder which will receive contributions both from $ \VHard $ and $ \VSoft $. In this way we can isolate the contributions that are purely due to hard modes. The remaining hard/soft modes can be dealt with separately.

Before the $ \hbar $-expansion is performed, an expansion is performed around $ \pbh $,
\begin{align*}
V(\pbmin)&=V(\pbh)+\pbs \left.\partial V\right|_{\pbh}+\frac{1}{2}\pbs^{2}\left.\partial^{2} V\right|_{\pbh}+\mathellipsis\\
&=\left\{\VHard(\pbh)\right\}+\VSoft(\pbh)+\pbs \left.\partial \VSoft\right|_{\pbh}\\
&+\frac{1}{2}\pbs^{2}\left.\left(\partial^{2} \VHard+\partial^{2} \VSoft\right)\right|_{\pbh}+\mathellipsis
\end{align*}
In this expansion we have isolated the purely hard part, $ \VHard(\pbh) $, within curly brackets. This contribution is denoted as $ \Vminh $, and is given by
\begin{align*}
\Vminh=\VHard_{0}(\pbh_{0})+&\hbar \VHard_{1}(\pbh_{0})\\
+&\hbar^{2}\left.\left(\VHard_{2}-\frac{1}{2}\left(\pbh_{1}\right)^{2}\partial^{2}\VHard_{0}\right)\right|_{\pbh_{0}}+\mathellipsis
\end{align*}
The purely soft terms and the soft-hard mix terms (also referred to as soft) are denoted by $ \Vmins $ and are given by
\begin{align*}
\Vmins=&\hbar \VSoft_{1}(\pbh_{0})\\
+&\hbar^{2}\left.\left(\VSoft_{2}-\frac{1}{2}\left(\pbs_{1}^{2}+2\pbh_{1}\pbs_{1}\right)\partial^{2}\VHard_{0}\right)\right|_{\pbh_{0}}+\mathellipsis
\end{align*}
Note that the leading order contribution to $ \VHard $ is $ \VHard_{0}=V_{0} $, and that the soft potential $ \VSoft $ starts at $ \VSoft_{1} $. By labeling $ V_{0} $ as hard, we are ensuring that the shifted Goldstone mass can be written as $ \overline{G}=\partial \VHard /\pb $. That is, we identify $ \pbh_{0}=\pb_{0},\pbs_{0}=0 $.

After isolating the hard and soft contributions to $ \Vmin $, we are ready to address the questions of gauge invariance and \ir{} fitness of the $ \hbar $-expansion. Ideally, we would like that the gauge dependence of the effective potential separates according to the hard/soft split. We perform and discuss this split in the next subsubsection. The main question is whether the soft contribution $ \Vmins $ is \ir{} finite. Even though we are not currently able to establish the \ir{} fitness of $\Vmins$ to all orders in perturbation theory, we have shown that it holds at least up to three loops. Additionally, leading divergences and a class of subleading divergences cancel to every loop order, see appendix \ref{sec:LDS}.
\subsubsection{Gauge dependence of $V^{\text{H}}$}\label{sssec:nielsen}
In order to assess the gauge dependence of $ \VHard $ we begin by considering the gauge dependence of the full effective potential, which can be studied through the Nielsen identity\room\cite{Nielsen:1975fs},
\begin{equation}\label{eq:NielsenIdent}
\left(\xi \partial_{\xi}+ C(\phi,\xi)\partial_\phi\right)V(\phi,\xi)=0,
\end{equation}
where $ C(\phi,\xi) $ is the Nielsen coefficient. In Fermi gauge Abelian Higgs, $ C(\phi,\xi) $ is given by
\begin{equation}\label{eq:NielsenCoeff}
	C(\pb,\xi)=
	\begin{gathered}
		\includegraphics[height=30pt]{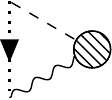}
	\end{gathered}=\frac{e}{2}\momint{k} \frac{1}{k^2} \left(-k_\mu \right) \overline{F}^\mu(k),
\end{equation}
where $ \overline{F}^{\mu} $ is the exact $G$-$A^\mu$ mixed propagator. The Nielsen identity, taken at face value, is a non-perturbative statement that implies that the effective potential is gauge invariant when evaluated at its extrema. However, in perturbation theory the situation is more subtle. The authors of\room\cite{Andreassen:2014eha} point out that $ C(\pb,\xi) $ can be perturbatively \ir{} divergent when evaluated at $ \pbmin $, which jeopardizes the conclusion that the effective potential is gauge invariant there. On the other hand, the authors of\room\cite{Patel:2011th} and the authors of\room\cite{Espinosa:2016uaw} argue that equation\room\eqref{eq:NielsenIdent} should be considered perturbatively. In performing an $\hbar$-expansion of the Nielsen identity, any \ir{} divergences of $ C(\pb,\xi) $ will cancel against zeros of $ \partial V $. Some of these subtleties are illustrated below, but we note that in the issue that we actually are interested in --- the gauge dependence of $ \VHard $ --- these concerns turn out to play no role.

The Nielsen coefficient $ C(\pb, \xi) $ describes the gauge dependence of $ V $. By the method of characteristics, it can be shown that curves $ \pb(\xi) $ in the $ \left(\pb,\xi\right) $ plane are curves of constant $ V $ if the curves satisfy
\begin{equation}\label{eq:characteristic}
\xi \partial_{\xi} \pb = C(\pb,\xi).
\end{equation}
This is true for the vev $ \pbmin(\xi) $.  In order to find a useful expression for $  C(\pb,\xi) $ we use the generalized Ward identities in appendix\room\ref{app:ward}. We find
\begin{align}
\overline{F}^\mu(k)&= \xi \sqrt{A}\frac{-k^\mu}{\left(k^{2}-\widetilde{G}_{+}\right)\left(k^{2}-\widetilde{G}_{-}\right)},\label{eq:exactMixedProp}\\
\widetilde{G}_\pm&\define\frac{1}{2}\left(\frac{\Gbar}{\Pi_{\textcaps{L}}}\pm\sqrt{\frac{\Gbar}{\Pi_{\textcaps{L}}}(\frac{\Gbar}{\Pi_{\textcaps{L}}}-4 \xi A)}\right),
\end{align}
where $ \Gbar = \frac{1}{\pb} \partial V $ and $ \Pi_{\textcaps{L}}(k^2) = 1+\ordo{}(\hbar) $ is a scalar function that parametrizes the self-energy of the longitudinal part of the $ (A^{\mu},G) $ matrix propagator. From equation\room\eqref{eq:NielsenCoeff} it can be seen that $ C(\pb,\xi) $ has an \ir{} divergence as $ \Gbar \rightarrow 0 $. The 1-loop result is the leading contribution and is previously known (see e.g.\room\cite{Espinosa:2016uaw}),
\begin{equation*}
C_{1}(\pb,\xi)= \frac{\xi A}{2\pb} \frac{1}{ \left(G_{+}-G_{-}\right)}\left[G_{+}\left(\LOG{G_{+}}-1\right)-G_{-}\left(\LOG{G_{-}}-1\right)\right].
\end{equation*}
The 1-loop Nielsen coefficient $C_{1}(\pb,\xi)$ is divergent in the $ \pb \rightarrow \pbmin_{0} $ limit. Interestingly, this divergence induces a $ \xi $ dependent \ir{} divergence in $ \pbmin_{1}(\xi) $ through equation\room\eqref{eq:characteristic}. This \ir{} divergence is not problematic because it is necessary to cancel other \ir{} divergences in the $ \hbar $-expansion.

Consider now the purely hard contribution. The Nielsen identity for the hard potential is
\begin{equation}\label{eq:NielsenHard}
\left(\xi \partial_{\xi}+ C^{\textcaps{h}}(\pb,\xi)\partial\right)\VHard(\pb,\xi)=0.
\end{equation}
The Nielsen coefficient $ C^{\textcaps{h}} $ can be found by using the mixed propagator given in equation\room\eqref{eq:exactMixedProp} but with the shifted Goldstone mass given by $\Gbar=\partial \VHard/\pb$. Note that the restriction to hard fields assumes $\smash{k^2 \gg \overline{G}_{\pm}}$. Hence, close to the minimum $ \pbh $ of the hard potential, we can expand
\begin{equation*}
\frac{1}{\left(k^{2}-\widetilde{G}_{+}\right)\left(k^{2}-\widetilde{G}_{-}\right)}=\frac{1}{k^4}+\ordo{}\left(\Gbar\right),
\end{equation*}
because by assumption $k^2\gg \overline{G}$.\footnote{Because we are only considering hard fields, the "hard" self energy $ \partial V^H/\pb$ is free of any divergences, which justifies our expansion.} In this momentum region the Nielsen coefficient behaves as
\begin{align*}
C^{\textcaps{h}}(\phi,\xi)|_{\phi\approx \pbh} =\frac{e}{2}\int(dk) k^{2}\frac{\xi \sqrt{A}}{k^4}+\mathcal{O}\left(\overline{G}\right)=0+\mathcal{O}\left(\overline{G}\right),
\end{align*}
where the last equality follows from the integral being scaleless. In particular, at $ \pb = \pbh $, we find $  C^{\textcaps{h}}(\pbh,\xi)=0 $. This is in sharp contrast to the behaviour of $ C(\pb,\xi)$ for the whole potential --- gone are the ambiguities associated with infinities.

It is not surprising that there are no \ir{} divergences for the hard fields because these can only come from long distance behaviour. More interesting is the fact that $  C^{\textcaps{h}}(\pbh,\xi)=0 $. First of all, this implies that the hard potential evaluated at $ \pbh $ is gauge invariant. Second of all, equation\room\eqref{eq:characteristic} implies that $ \pbh $ is gauge invariant. In other words, $ \pb=\pbh $ is an invariant \textcaps{1d} manifold in the $ \left(\pb, \xi\right) $ plane under the flow defined by the Nielsen coefficient through equation\room\eqref{eq:characteristic}. It is also understood that $ \pbh $ does not have any \ir{} divergences.

To summarize, we have now shown that
\begin{itemize}
	\item $ \VHard $ is gauge invariant when evaluated at $ \pbh $. This means that the gauge dependence of $ \VHard $ and $ \VSoft $ separates.
	\item $ \pbh $ is gauge invariant. Note that this places heavy restrictions on the gauge dependence of $ \Gbar = G + \Delta $, because $ \pbh $ is defined through $ \left.\Gbar\right|_{\pbh}=0 $.
\end{itemize}
\subsubsection{IR fitness of the $ \hbar $-expansion}\label{sssec:irfit}
Because $ \Vminh $ is finite and gauge invariant, we address the \ir{} finiteness of $ \Vmins $. Even if the soft effective potential $ \VSoft $ receives \ir{} divergent contributions, it is not guaranteed that $  \Vmins $ is \ir{} divergent. Gauge dependent divergences are guaranteed to cancel order by order in $\hbar$. This section focuses on the "gauge independent" divergences found in Landau gauge ($\xi=0$), and show how these cancel.

The idea is that divergences in the $\hbar$-expansion of $ \Vmins $,
\begin{align*}
\Vmins=&\hbar \VSoft_{1}(\pb_{0})\\
+&\hbar^{2}\left.\left(\VSoft_{2}-\frac{1}{2}\left(\pbs_{1}^{2}+2\pbh_{1}\pbs_{1}\right)\partial^{2}V_{0}\right)\right|_{\pb_{0}}+\mathellipsis,
\end{align*}
might cancel order by order. We note that in Landau gauge, insertions of $ \pbs_{1} \propto G $ will soften any \ir{} divergence.

\vspace{\leadingHeight}
\noindent\textbf{3-loop singularity cancellation:}\quad To illustrate how such a cancellation can work, let us consider the logarithmic ($\sim \log G$) divergence at three loops. In the following we will only retain possibly divergent terms; we use the symbol $ \simeq $ to denote expressions equivalent up to finite terms. The $\hbar$-expansion of $ \Vmins $, with all terms implicitly evaluated at $\phi_0$, gives
\begin{align}\label{eq:3loop}
\begin{split}
\ordo{}(\hbar^3):\qquad &\left\{\VSoft_{3}+\pbh_{1}\partial \VSoft_{2}+\frac{\left(\pbh_{1}\right)^2}{2!}\partial^2 \VSoft_{1}\right\}\\
+&\left[\pbs_{2}\left(\partial \VSoft_{1}+\pbh_{1} \partial^2 V_0\right) \right].
\end{split}
\end{align}
The possible divergences of the expression within the curly brackets in equation\room\eqref{eq:3loop} come from the 2- and 3-loop daisy diagrams
\begin{align*}
	\VSoft_2\sim
	\begin{gathered}
		\includegraphics[height=30pt]{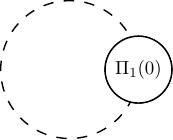}
	\end{gathered}\room,~\VSoft_3\sim
	\begin{gathered}
		\includegraphics[height=30pt]{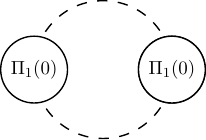}
	\end{gathered}\room,
\end{align*}
where the Goldstone self-energy insertions are hard. Our aim is now to translate these diagrams to the language of the $ \hbar $-expansion. For a general $ l $-loop daisy diagram, the contribution to the potential is
\begin{align}\label{eq:ldaisy}
\VSoft_{l}=-\frac{1}{2}\frac{(\Pi^{\textcaps{h}}_1(0))^{l-1}}{l-1}\momint{k} \frac{(-1)^{l-1}}{(k^2+G)^{l-1}}.
\end{align}
We can relate this expression to the soft 1-loop potential via the relation
\begin{equation*}
-\frac{1}{2}\momint{k}\frac{1}{(k^2+G)^n}=\frac{(-1)^n\partial_{G}^n \VSoft_{1}}{(n-1)!},
\end{equation*}
which gives
\begin{equation}\label{eq:ldaisyDeriv}
\VSoft_{l}=\frac{1}{\left(l-1\right)!}\left(\Pi_{1}^{\textcaps{h}}(0)\right)^{l-1}\partial_{G}^{l-1} \VSoft_{1}.
\end{equation}
The \ir{} divergent contributions to the 2- and 3-loop $ \VSoft $ can then be written
\begin{align*}
& \VSoft_2 \simeq (\Pi^{\textcaps{h}}_1(0)) \partial_{G} V^S_1,
\\& \VSoft_3 \simeq (\Pi^{\textcaps{h}}_1(0))^2 \frac{\partial_{G}^2 V^S_1}{2}.
\end{align*}
By using the definition of the Goldstone mass, the derivatives with respect to G can be rewritten as
\begin{align*}
&\partial_{G}=\frac{3}{\lambda \phi}\partial,
\\& \partial_{G}^2=\left(\frac{3}{\lambda \phi}\right)^2\left(-\frac{1}{\phi}\partial+\partial^2\right).
\end{align*}
We give a general formula for an arbitrary number of $ G $ derivatives in appendix\room\ref{ss:subleading}.

When $ \pb $-derivatives act on the soft 1-loop potential, the result scales as $\partial^n \VSoft_{1}\sim G^{2-n}$. Thus the logarithmically divergent terms that are present at three loops can only come from $\partial^2 V_1^S$. We now use $\Pi^{\textcaps{h}}_{1}(0)=-\partial^2 V_0 \pbh_{1}/\pb_{0}$ and $\partial^2 V_0=\lambda\phi_0^2/3$ to rewrite the divergent terms in the $\hbar$-expansion as
\begin{align*}
&\VSoft_{3}+\pbh_{1}\partial \VSoft_2+\frac{\left(\pbh_{1}\right)^2}{2}\partial^2 \VSoft_1 \simeq\\
&(\pbh_{1})^2\partial^2 \VSoft_1\left(\frac{1}{2}-1+\frac{1}{2}\right)=0.
\end{align*}

The remaining terms in the straight brackets in equation\room\eqref{eq:3loop} are possibly divergent because they are proportional to $\pbs_{2}$; we have not established whether this quantity is finite yet. It is defined by
\begin{align*}
\hbar^2\left(\partial^2 V_0 \pbs_{2}+\partial \VSoft_2+\pbh_{1} \partial^2 \VSoft_{1}\right)\simeq 0.
\end{align*}
The divergent contributions to $ \pbs_{2} $ are
\begin{align*}
\pbs_{2}&\simeq-\frac{1}{\partial^{2}V_{0}}\left( \partial \VSoft_2+\pbh_{1} \partial^2 \VSoft\right)\\
&\simeq-(\pbh_{1})^2 \frac{\partial^2 \VSoft_1}{\partial^2 V_{0}}\left(-1+1\right)=0,
\end{align*}
and  $\pbs_{2}$ is finite. We have showed that $ \Vmins $ is finite (actually zero) to three loops in the $ \hbar $-expansion.

\vspace{\leadingHeight}
\noindent\textbf{Leading singularity cancellation:}\quad The above example can be generalized. The idea is that we can rewrite any daisy diagram in terms of derivatives acting on $ \VSoft_{1} $. We will use this procedure to show that leading singularities cancel to all orders, where leading singularities are defined as the worst possible divergence at a given loop order. In this calculation we use the symbol $ \simeq $ to denote equivalence up to sub-leading divergences (note that this definition carries over to the three loop example).

At $ L $ loops the leading singularity behaves as $V_L\sim G^{3-L}$. At the $\hbar^L$ order, the $\hbar$-expansion gives terms of the form
\begin{align*}
\Vmins=\ldots+\hbar^L\left(\VSoft_L+\phi_{1}\partial \VSoft_{L-1}+\ldots+\frac{\phi_{1}^{L-1} \partial^{L-1}\VSoft_1}{(L-1)!}\right)+\ldots
\end{align*}
If $L\geq 3$ all the terms in the parenthesis have the same leading singularity. This can be seen from that the leading singularity for $\partial^n V_l$ is $\sim G^{3-l-n}$, and $l+n=L$ for all terms in the parenthesis. The $ \hbar^{L} $ leading singularity contribution can hence be written as
\begin{equation}\label{eq:VminSL}
\left.\Vmins\right|_{L}\simeq \sum_{n=0}^{L-1}\frac{\phi_1^{n}\partial^n \VSoft_{L-n}}{n!}.
\end{equation}
Again, our goal is to rewrite this in terms of derivatives acting on $ \VSoft_{1} $. The expression for the $ l $:th loop level daisy is given in equation\room\eqref{eq:ldaisyDeriv}, in terms of $ G $-derivatives acting on $ \VSoft_{1} $. To exchange these derivatives for $ \pb $-derivatives we note that when counting leading singularities,
\begin{equation}
\partial_{G}^{n}\simeq\left(\frac{3}{\lambda \pb}\right)^{n}\partial^{n}.
\end{equation}
We can express $ \VSoft_{l} $ as
\begin{align*}
\VSoft_l\simeq \frac{\phi_1^{l-1}(-1)^{l-1}\partial^{l-1}\VSoft_1}{(l-1)!},
\end{align*}
which allows us to perform the sum in equation\room\eqref{eq:VminSL},
\begin{align*}
&\sum_{n=0}^{L-1}\frac{\phi_1^{n}\partial^n \VSoft_{L-n}}{n!}\simeq\\
&\phi_1^{L-1}(-1)^{L-1}\partial^{L-1} \VSoft_1 \sum_{n=0}^{L-1}\frac{(-1)^n}{n!(L-1-n)!}=0.
\end{align*}
Here we made use of the Binomial identity
\begin{align}\label{eq:Binomial}
\sum_{n=0}^c (-1)^n\binom{c}{n}=0.
\end{align}
We see that the leading singularities cancel order by order in $\hbar$.

In appendix \ref{sec:LDS} we extend the machinery developed here to show the even stronger result that the leading singularities cancel if $\Pi_1(0)\rightarrow \Pi(k^2)$. The reader might worry about divergences in $\phi_L$. However, the leading singularities in $\phi_L$ are again proportional to the sum $\sum_{n=0}^{L-1}\frac{1}{n!}\phi_1^{n}\partial^n \VSoft_{L-n}$, and vanish.

What about sub-divergences? The complete fitness of $V_{\text{min}}$ has not yet been shown, but we have shown that a particular class of sub-divergences cancel. The combinatorics of this cancellation is non-trivial, as we show in appendix\room\ref{sec:LDS}. On this basis, we conjecture that $\Vmins=0$ holds to all loop orders, with the full proof left for the future.
\subsection{Quantitative results}\label{ssec:quantitative}
Having introduced our strategy for isolating the physical part of $ \Vmin $ in subsection\room\ref{ssec:strategy}, we now turn to demonstrating it by calculations in both the Abelian Higgs and the \cw{} model.
\subsubsection{Abelian Higgs}\label{sssec:AH}
The Abelian Higgs model is the simplest model to illustrate the strategy discussed in the previous subsections. In Fermi gauges the 1-loop contribution is given by
\begin{align*}
V_{1}\left(\pb\right)=&\frac{1}{4}\left[H^{2}\left(\LOG{H}-\frac{3}{2}\right)+3 A^{2}\left(\LOG{A}-\frac{5}{6}\right)\right.\\
+&\left. G_{+}^{2}\left(\LOG{G_{+}}-\frac{3}{2}\right)
+ G_{-}^{2}\left(\LOG{G_{-}}-\frac{3}{2}\right)\right].
\end{align*}
The hard and soft contributions can be quickly identified: $ \VHard_{1} $ is given by the contributions from $ H $ and $ A $, while $ \VSoft_{1} $ is given by $ G_{+} $ and $ G_{-} $. With these 1-loop contributions we can find $ \pbmin_{1}$ through the procedure delineated below equation\room\eqref{eq:diffV}. The definition of $ \pbh $ is parallel to that of $ \pbmin $, and it is hence straightforward to perform the separation $ \pbmin_{1}=\pbh_{1}+\pbs_{1} $. The result is
\begin{align*}
\pbh_{1}=&-\sqrt{\frac{3 m^{2}}{2 \lambda^{3}}} \left(18 e^4 \left(\log\left[\frac{6 e^2 m^{2}/\lambda}{\mu^{2}}\right]-\frac{1}{3}\right)\right.\\
&+\left.\lambda^2\left(\log\left[\frac{2 m^{2}}{\mu^{2}}\right]-1\right)\right),\\
\pbs_{1}=&e^2 m^2 \xi \frac{1}{2}\sqrt{\frac{3 m^2}{2\lambda}}\left(\LOG{G}+\log\left[\frac{6 \xi e^2 m^2  /\lambda}{\mu^2}\right]-2\right).
\end{align*}
Note that $\pbs_1$ has an IR divergence of the form $\varphi|_{G\rightarrow 0} \sim \xi \LOG{G}$.

For the 2-loop effective potential calculation we are interested in separating it into its hard and soft parts, $ V_{2}(\pb)=\VHard_{2}(\pb)+\VSoft_{2}(\pb) $. In practice this amounts to using the method of regions to separate the momentum integrals into integrals over hard and soft modes, as described in\room\cite{Espinosa:2017aew}. It should also be noted that the method of regions can be applied directly to the renormalized momentum integrals, to ensure that any finite contribution from counterterm insertions get identified correctly. We use renormalized integrals throughout this section.

At two loops we expect \ir{} divergences from the soft contributions. The outline of the calculation of the 2-loop potential is given in appendix\room\ref{app:2loop}. Before we consider the full soft contribution, let us consider the double Goldstone bubble diagram $ \mathcal{M}_{\text{B}} $. This diagram is an all-lines-soft diagram that is \ir{} divergent. Its value at $ \pb=\pbmin_{0} $ is given by
\begin{align*}
	\left.\mathcal{M}_{\text{B}}\right|_{\pbmin_{0}} &=
	\left.\begin{gathered}
		\includegraphics[height=25pt]{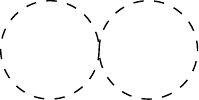}
	\end{gathered}\right|_{\pbmin_{0}} \!\!= \left.-i^2\lambda\frac{1}{2^3}\left(\momint{q}D_{G}(q)\right)^{2}\right|_{\pbmin_{0}}\\
	&=\frac{1}{32}\frac{e^{4}m^{4}}{\lambda}\xi^{2}\left(\LOG{G}+\log\left[\frac{6 \xi e^{2} m^{2}/\lambda}{\mu^{2}}\right]-2\right)^{2}.
\end{align*}
As we discussed in subsection\room\ref{ssec:powercountingfermi}, this diagram is an example of the kind of \ir{} divergence that prevents us from performing a resummation in Fermi gauges. Additionally, note the presence of the $\LOG{G}^{2}$ term, which is not possible to generate in Landau gauge at two loops.

At the tree level minimum $\pbmin_{0}$, the 2-loop level soft potential is given by

\begin{widetext}
\small
\begin{align*}
\left.\VSoft_{2}\right|_{\pbmin_{0}}&= \frac{3 e^2 m^4 \xi}{4\lambda^2}\left\{ \frac{e^2 \lambda \xi}{2}\LOG{G}^2\right.-\LOG{G}\left(-12 e^4+2e^2 \lambda\xi-e^2\lambda\xi\log\left[\frac{6\xi e^2 m^2 /\lambda}{ \mu^2}\right]\right.\left.+36 e^4 \log\left[\frac{6e^2 m^2/\lambda}{\mu^2}
\right]
-2 \lambda^2+2\lambda^2 \log\left[\frac{2 m^2}{\mu^2}\right]\right)\\
&-\frac{1}{2}\left(\log\left[\frac{6 e^{2}m^{2}\xi/\lambda}{\mu^{2}}\right]-2\right)\left[72 e^{4}\left(\log\left[
\frac{6 e^{2}m^{2} / \lambda}{\mu^{2}}\right]-\frac{1}{3}\right)\right.
\left.\left.+4\lambda^{2}\left(\log \left[\frac{2 m^{2}}{\mu^{2}}\right]-1\right)-e^{2}\lambda\xi \left(\log\left[\frac{6 e^{2}m^{2}\xi/\lambda}{\mu^{2}}\right]-2\right)\right]\right\}.
\end{align*}
\end{widetext}

The full $ \hbar^{2} $ contribution to $ \Vmins $ is given by
\begin{align*}
\left.\left(\VSoft_{2}-\frac{1}{2}\left(\pbs_{1}^{2}+2\pbh\pbs\right)\partial^{2}V_{0}\right)\right|_{\pbmin_{0}}.
\end{align*}

After some straightforward algebra one finds that $ \Vmins $ is \ir{} finite, and in fact zero, as promised.

\begin{figure}
	\begin{centering}
	\includegraphics[width=\linewidth]{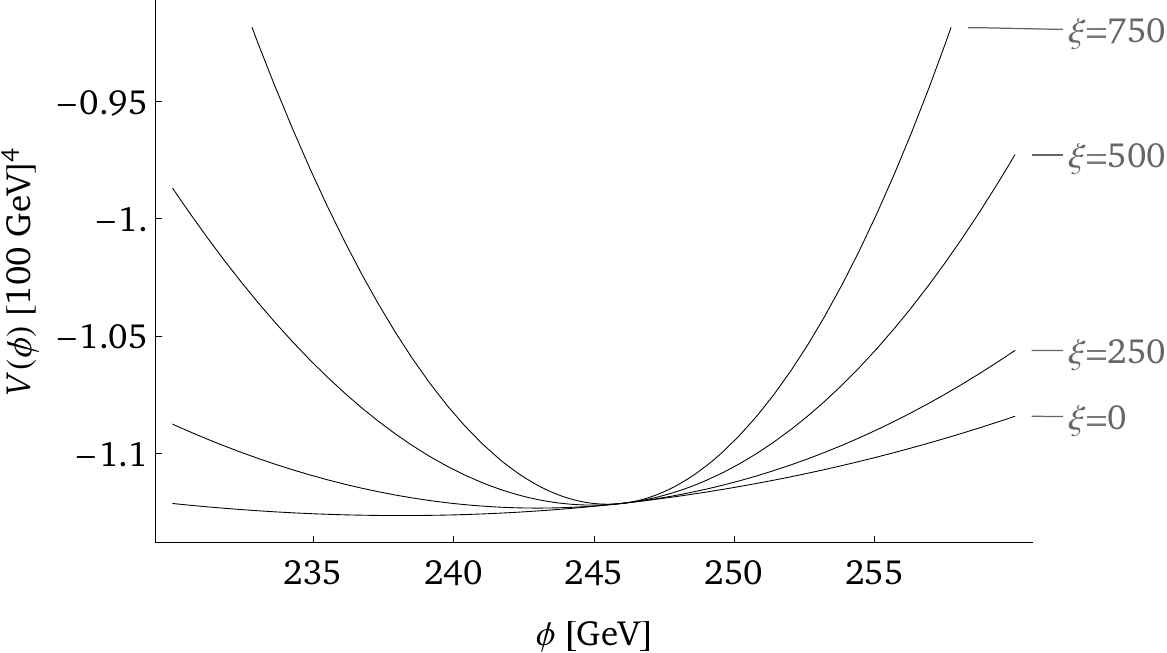}
	\caption{The hard part of the 2-loop effective potential, $ \VHard(\pb) = V_{0}(\pb)+\hbar \VHard_{1}(\pb)+\hbar^{2} \VHard_{2}(\pb)$, versus the background field $ \pb $. The parameters are chosen such that $ \pbmin_{0} = 246\room\text{GeV},\room \left. H\right|_{\pbmin_{0}}=\left(125\room\text{GeV}\right)^{2},\room\left. A\right|_{\pbmin_{0}}=\left(90\room\text{GeV}\right)^{2} $.\label{fig:V2hard}}
	\end{centering}
\end{figure}

We do not give the full 2-loop contribution to the hard effective potential, because its full form is not particularly enlightening. It does however have a few interesting properties. We find that the gauge dependence of $ \VHard_{2}(\pb) $ separates according to $ \VHard_{2}(\pb)=\left.\VHard_{2}(\pb)\right|_{G=0}+f(\xi,\pb)G^{2}+\ordo{}(G^{3})  $, where $ f(\xi,\pb) $ is a second degree polynomial in $ \xi $. This behavior is illustrated in figure\room\ref{fig:V2hard}, where we plot the 2-loop hard potential for a benchmark parameter point. Even though the cancellation of $ \Vmins$ suggests that the physical content of $ \Vmin $ is contained in $ \Vminh $, minimizing $ \VHard $ numerically will still yield a gauge dependent result (though we note that the result converges to $ \VHard (\pbmin_{0}) $ in the limit $ \xi \rightarrow \infty $). We also note that the hard Goldstone self-energy, $ \Delta=\partial \VHard / \pb $, will in general depend on $ \xi $.

The fact that the gauge dependent terms of $ \VHard_{2}(\pb) $ scale with $ G^{2} $ ensures that the potential and its first derivative are gauge invariant when evaluated at $ \pbmin_{0} $. Consequently, this guarantees the gauge invariance of the $ \hbar^{2} $ contributions to $ \Vminh $,
\begin{align*}
\Vminh=\left.V_0\right|_{\pbmin_0}+&\hbar\left.\VHard_1\right|_{\pbmin_0}\\
+&\hbar^2\left.\left(\VHard_{2}-\frac{1}{2}(\pbh_{1})^2 \partial^2 V_0\right)\right|_{\pbmin_0}+\mathellipsis,
\end{align*}
and $ \pbh $, in accordance with our analysis of the Nielsen identity in subsubsection\room\ref{sssec:nielsen}. Even with this in mind, it is easier to calculate the hard potential compared to calculating the full potential and then ensuring cancellations of the \ir{} divergences.

\vspace{\leadingHeight}
\noindent \textbf{RG invariance:}\quad
Physical quantities should be \rg{} invariant to ensure that the results do not depend on the fictional scale used. In this way, \rg{} invariance serves as an important check of the results.

The 2-loop potential $V=V_0+V_1+V_2+\mathcal{O}\left(\hbar^3\right)$ satisfies the \rg{} equation

\vspace{-.5em}
\begingroup
\footnotesize
\begin{align}\label{eq:RG}
\left(\mu \frac{\partial}{\partial\mu}-\phi \gamma_{\pb} \frac{\partial}{\partial\phi}+\beta_\lambda \frac{\partial}{\partial\lambda}+\beta_e \frac{\partial}{\partial e}+\beta_{m^2}\frac{\partial}{\partial m^2}+\beta_\xi \frac{\partial}{\partial\xi}\right)V=0.
\end{align}
\endgroup%
We want to check whether this equation is satisfied up to $ \ordo{}(\hbar^{2}). $ Because the parameters $ e  $ and $ \xi $ first contribute at the 1-loop level, it is sufficient to retain the leading order $\beta $-functions for them, both of which are well known. They are given by
\begin{align*}
&\beta_e=\hbar \frac{e^3}{3}+\mathcal{O}\left(\hbar^2\right),
\\&\beta_\xi=-\hbar \xi \frac{2 e^2}{3}+\mathcal{O}\left(\hbar^2\right).
\end{align*}
The gauge parameter beta function $\beta_\xi$ is absent in many standard calculations because Landau gauge is the common gauge choice for calculating the effective potential in which $\beta_\xi=0$.

The anomalous dimension and remaining beta functions are given by\footnote{The 2-loop beta function for the mass $\beta_{m^2}$ is not explicitly given in the literature, but as noted in \cite{Martin:1993zk} it can easily be extracted by introducing a dummy field , $\phi_3$, and rewriting the mass term  $\frac{1}{2}m^2\left(\phi_1^2+\phi_2^2\right)=\frac{1}{4!}\lambda_{abcd} \phi_a \phi_b \phi_c \phi_d$, with $a,b,c,d=1,2,3$\room\cite{Machacek:1984zw,Machacek:1983tz}.}
\begingroup
\allowdisplaybreaks
\begin{align*}
\gamma_\phi=&\hbar e^2(\xi -3)+\hbar^2\left(\frac{10}{3}e^4+\frac{\lambda^2}{9}\right)+\mathcal{O}\left(\hbar^3\right),
\\\beta_\lambda=&\hbar\left(36 e^4-12e^2\lambda+\frac{10}{3}\lambda^2\right)
\\+&\hbar^2\!\left(-416 e^6\!+\!\frac{316}{3}e^4 \lambda\!+\!\frac{56}{3}e^2 \lambda^2\!-\!\frac{20}{3}\lambda^3\right)\!+\!\mathcal{O}\left(\hbar^3\right),
\\\beta_{m^2}= &\hbar m^2\left(\frac{4}{3}\lambda-6e^2\right)
\\+&\hbar^2 m^2\left(-\frac{10}{9}\lambda^2+\frac{32}{3}e^2 \lambda+\frac{86}{3}e^4\right)+\mathcal{O}\left(\hbar^3\right).
\end{align*}
\endgroup
With the above \rg{}-coefficients, the 2-loop effective potential $\left(V-V|_{\phi=0}\right)$ indeed fulfills the \rg{}-equation\room\eqref{eq:RG}, where $V|_{\phi=0}$ has been subtracted because renormalization of the zero-point vacuum energy is neglected.

In our strategy to calculate $ \Vmin $ we noted that the soft contributions cancel, and that the purely hard contribution $ \Vminh=\VHard(\pbh) $ is all that remains. From our calculation above, we expect then that $ \Vminh $ is \rg{} invariant, because it is just the finite contribution to $ \Vmin $. We can confirm this explicitly by checking the \rg{} invariance of $ \Vminh $. This invariance takes the form
\begin{align}\label{eq:RGH}
\left(\mu \frac{\partial}{\partial\mu}+\beta_\lambda \frac{\partial}{\partial\lambda}+\beta_e \frac{\partial}{\partial e}+\beta_{m^2}\frac{\partial}{\partial m^2}\right)\VHard|_{\pb=\pbh}=0.
\end{align}
Note that in this \rg{} eqation we could neglect the contributions from rescaling the field with $ \gamma_\phi $ and the evolution of $ \xi $ through $ \beta_\xi $, because $ \VHard $ is evaluated at its extremum $ \pbh $, and because we have shown that it is $ \xi $-independent there.

It is straightforward to check that $\VHard$ fulfills the \rg{} equation\room\eqref{eq:RGH} at leading order. At the $\mathcal{O}\left(\hbar^2 \right)$ level the \rg{} equation assumes the form
\begin{align*}
&\big{\{} \mu\frac{\partial}{\partial\mu} \VHard_{2}+\pbh_1 \mu\frac{\partial}{\partial\mu}\partial_{\pb} \VHard_{1}+\left(\beta^1_{m^2} \partial_{m^2}+\beta^1_{\lambda} \partial_\lambda+\beta^1_{e} \partial_e\right)\VHard_{1}
\\+&(\beta^2_\lambda \partial_\lambda +\beta^2_{m^2} \partial_{m^2}+\pbh_{1} \beta^1_{m^2} \partial_{m^2}+\pbh_1 \beta^1_{\lambda} \partial_\lambda)\VHard_0\big{\}}\big{\}}_{\phi=\phi_0}=0.
\end{align*}
By explicit calculation $\VHard_{2}|_{\phi=\phi^H}-\VHard_{2}|_{\phi=0}$ is indeed \rg{} invariant to $\mathcal{O}\left(\hbar^2\right)$. We expect that this holds to all orders in $ \hbar $, because we have explicitly separated the hard and soft degrees of freedom. The hard and soft quantities should be individually \rg{} invariant because they do not "talk" with each other.

To summarize, we have seen that $ \VHard $ is \rg{} invariant when evaluated at $ \pbh $ up to $ \ordo{}(\hbar^{2}) $. This means that the $\mu$ dependence of $ \VHard $ and $ \VSoft $ separates, at least up to $ \ordo{}(\hbar^{2}) $, but presumably to all orders. This supports the claim that $\Vminh$ is a physically meaningful quantity.
\subsubsection{The \textcaps{CW} model}\label{sssec:cw}
The \cw{} model is a special instance of scalar \qed{} mentioned in the introduction. This model is defined by $ m^{2}=0 $ and does not have spontaneous symmetry breaking at the classical level, but it can exhibit symmetry breaking at the quantum level. S. Coleman and E. Weinberg emphasized in their original paper\room\cite{Coleman} that this requires a careful treatment of perturbation theory, with the coupling $ \lambda $ scaling as $ \lambda \sim \hbar $. Recently Andreasson, Frost, and Schwartz computed the effective potential in the \cw{} model for the Fermi gauges up to two loops and demonstrated that the scaling $ \lambda\sim \hbar $ means that it is no longer enough to just perform the $ \hbar $-expansion to establish gauge invariance of $ \Vmin $\room\cite{Andreassen:2014eha}. There are daisy diagrams which contribute as $ \frac{1}{\lambda^{l}} $ at loop level $ l $, which breaks the $ \hbar $ power counting. This necessitates a resummation of these terms in order to cure the gauge dependence. Later it was pointed out by Espinosa, Garny, and Konstandin\room\cite{Espinosa:2016uaw} that this resummation is the same as the one that two of the authors developed to cure the \ir{} issues of the effective potential in\room\cite{Elias-Miro:2014pca}. Here we want to clarify this relation by being very explicit about the hard/soft separation of $ \Vmin $ in the \cw{} model, and to place this discussion in the context of the $ \hbar $-expansion. We also discuss our interpretation of why perturbation theory breaks down in this model.

The masses in this model are
\begin{align*}
G&=\frac{1}{6}\lambda\pb^{2},\\
H&=\frac{1}{2}\lambda\pb^{2}=3 G,\\
A&=e^{2}\pb^{2}.
\end{align*}
Even though the $ H $ field is on the same footing as $ G $, we choose to still retain the original labels. The mixing masses $ G_{\pm} $ are given in terms of the above masses just as in the Abelian Higgs model. We implement the $ \lambda \sim \hbar $ scaling by extracting a factor of $ \hbar $, i.e. we let $ \lambda \rightarrow \hbar \lambda $. This pushes each factor of $ \lambda $ up one order in the perturbation expansion, and we hence have that $ V_{0}(\pb)=0 $. The 1-loop potential is
$ V_{1}(\pb)=\frac{\lambda}{4!}\pb^{4} + \frac{3}{4}A^{2}(\LOG{A}-\frac{5}{6}), $ and we can notice that $ V_{1}(\pb)=\VHard_{1} $, $ \VSoft_{1}=0 $, i.e. the 1-loop potential is purely due to hard contributions. The \emph{a priori} 1-loop terms from $ H $ and
$ G_{\pm} $ are pushed to the 2-loop level, and will give a part of the 2-loop soft contribution. By minimizing $ \VHard_{1}(\pb) $ we can find $ \pbh_{0} $, which is the leading order contribution to $ \pbh $,
$ 	\smash{\pbh_{0}=\frac{1}{e}\mu \exp [\frac{1}{6}-\frac{\lambda}{36 e^{4}}]}. $ We note that $ \pbh_{0} $ actually scales as $ \hbar^{0} $, even though it is defined through the 1-loop potential. This allows us to treat it just as the tree-level \vev{} $ \pb_{0} $ in ordinary perturbation theory, and expand around it. Remarkably, this carries on to the higher orders:
$ \pbh_{1} $ scales as $ \hbar $ even though it is generated at the 2-loop level, and so on. We of course also extract the 1-loop corrected Goldstone mass,
$ \smash{\ordo{}(\hbar):\room\overline{G} = \frac{1}{\pb}\partial \VHard= G + \frac{3}{2} e^{2} A \left( \LOG{A}-\frac{1}{3}\right)}, $ and this is by definition zero when evaluated at $ \pbh_{0} $.

Turning to the 2-loop diagrams, in order to perform the split into hard and soft it is easiest to work with renormalized integrals, as described in\room\cite{Espinosa:2017aew}. Using the same notation as in\room\cite{Andreassen:2014eha}, the non-zero diagrams at the 2-loop level are
\begin{alignat*}{3}
	\mathcal{M}_{\text{A}} &= 	\begin{gathered}
		\includegraphics[height=30pt]{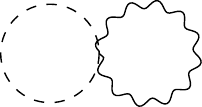}
	\end{gathered}\room, \qquad
	&&\mathcal{M}_{\text{B}} &&= 		\begin{gathered}
		\includegraphics[height=30pt]{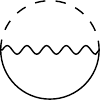}
	\end{gathered}\room,\\
	\mathcal{M}_{\text{C}}&=
	\begin{gathered}
		\includegraphics[height=30pt]{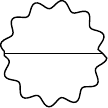}
	\end{gathered}\room, \qquad
	&&\mathcal{M}_{\text{D}} &&= \begin{gathered}
		\includegraphics[height=30pt]{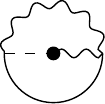}
	\end{gathered}\room.
\end{alignat*}

Evaluating these diagrams one finds (with terms $ \sim \lambda $ pushed to higher orders)
\begingroup
\allowdisplaybreaks
\begin{alignat*}{3}
\mathcal{M}_{\text{A}}&=\mathcal{M}_{\text{A}}^{\textcaps{s}}&&=
e^2 A^2 \xi&&\left[ -\frac{3}{4} \left( \LOG{A}-\frac{1}{3}\right) \left(\LOG{\xi A}+\LOG{G}-2\right)\right],\\
\mathcal{M}_{\text{B}}&= \mathcal{M}_{\text{B}}^{\textcaps{h}}&&=
e^2 A^2 &&\left[\frac{1}{4}\left(1+3\xi\right)L_{A}^{2}-\frac{1}{2}\left(2+4\xi\right)\LOG{A}\right.\\*
& && &&\left.+\frac{5}{4}+\frac{\pi^{2}}{12}+\frac{9}{4} \xi +\frac{\pi^{2}}{4}\xi\right],\\
\mathcal{M}_{\text{C}}&=\mathcal{M}_{\text{C}}^{\textcaps{h}}&&=
e^2 A^{2} &&\left[\frac{1}{4} \left(9+3\xi\right) \LOG{A}^2-\frac{1}{2}\left(\frac{56}{3}+4\xi\right) \LOG{A}\right.\\*
& && &&\left.+\frac{127}{12} -\frac{\pi ^2}{12}+\frac{9}{4} \xi +\frac{\pi^{2}}{4}\xi\right],\\
\mathcal{M}_{\text{D}}&=\mathcal{M}_{\text{D}}^{\textcaps{h}}&&=
e^2 A^{2}\xi &&\left[-\frac{3}{2}\LOG{A}^{2}+4\LOG{A}-\frac{9}{2}-\frac{\pi^2}{2}\right],
\end{alignat*}
\endgroup
where we noted that all diagrams are purely hard or soft. In\room\cite{Andreassen:2014eha} the authors note that they confirm through explicit calculation that $ \mathcal{M}_{\text{B}} +\mathcal{M}_{\text{C}}+\mathcal{M}_{\text{D}}$ is gauge invariant on its own, which was originally shown in\room\cite{Metaxas:1995ab}. We repeat this confirmation, and add that it goes hand in hand with the factorization into hard and soft. After adding these diagrams, the hard part of the 2-loop potential is
\begin{equation*}
\VHard_{2}(\pb)=\mathcal{M}_{\text{B}}^{\textcaps{h}}+\mathcal{M}_{\text{C}}^{\textcaps{h}}+\mathcal{M}_{\text{D}}^{\textcaps{h}}=\frac{1}{2}e^{2}A^{2}\left(5 \LOG{A}^{2}-\frac{62}{3}\LOG{A}+\frac{71}{3}\right).
\end{equation*}
We note that this expression reproduces the result in\room\cite{Espinosa:2016uaw}. Similarly, the soft part of the 2-loop potential is found by adding $ \mathcal{M}_{\text{A}}^{\textcaps{s}} $ to the terms pushed from the 1-loop level,
\begin{align*}
\VSoft_{2}(\pb)&=[\sim \lambda\text{ terms from } V_{1}(\pb)]+\mathcal{M}_{\text{A}}^{\textcaps{s}}\\
&=\frac{1}{4}H^{2}(\LOG{H}-\frac{3}{2})+\frac{1}{4} G_{+}^{2}(\LOG{G_{+}}-\frac{3}{2})+\frac{1}{4}G_{-}^{2}(\LOG{G_{-}}-\frac{3}{2})\\
&-\frac{3}{4}\xi e^{2}A^{2}\left(\LOG{A}-\frac{1}{3}\right)\left(\LOG{\xi A}+\LOG{G}-2\right).
\end{align*}
If we evaluate the 2-loop potential $ V_{2}=\VHard_{2}+\VSoft_{2} $ at $ \pbh_{0} $ we will find a gauge dependent result, as observed in\room\cite{Andreassen:2014eha}. Also, the contribution from $ \mathcal{M}_{\text{A}}^{\textcaps{s}} $ is divergent in the limit $ G \rightarrow 0 $ (though not in the limit
$ \pb \rightarrow \pbh_{0} $). We agree with the joint conclusion of references\room\cite{Andreassen:2014eha,Espinosa:2016uaw}, that both of these issues are solved through resummation. The process of resumming the daisies consists at this loop level of removing the one-petal daisy $ \mathcal{M}_{\text{A}}^{\textcaps{s}} $ and letting $ G \rightarrow \Gbar $ in the remaining "1-loop" terms, as described in subsubsection\room\ref{ssec:IR},
\begin{align*}
\overline{\VSoft_{2}}(\pb)&=\frac{1}{4}\overline{H}^{2}(\LOG{\overline{H}}-\frac{3}{2})\! + \! \frac{1}{4} \overline{G}_{+}^{2}(\LOG{\overline{G}_{+}}-\frac{3}{2})\! + \! \frac{1}{4}\overline{G}_{-}^{2}(\LOG{\overline{G}_{-}}-\frac{3}{2}),
\end{align*}
yielding a result that vanishes at $ \pbh_{0}  $ and hence is gauge invariant there. It is also finite in the limits $ G\rightarrow 0, \Gbar \rightarrow 0 $.

We can now evalulate $ \Vmin $ up to order $ \hbar^{2} $,
\begin{align*}
\Vmin=& \left\{\hbar\VHard_{1}(\pbh_{0})+\hbar^{2}\VHard_{2}(\pbh_{0})+\ordo\left(\hbar^{3}\right)\right\}+0\\
=&-\hbar\frac{3}{8}\exp{\left[\frac{2}{3}-\frac{\lambda }{9 e^4}\right]}\mu^{4}\\*
&+\hbar^{2}\exp{\left[\frac{2}{3}-\frac{\lambda }{9 e^4}\right]}\mu^{4} \left(\frac{26}{3} e^8+\frac{13}{27} e^4 \lambda \!+\!\frac{5}{648} \lambda ^2\right).
\end{align*}
We note that this is gauge invariant for arbitrary values of the \msbar{} scale $ \mu $. However, if we evaluate this at the scale $ \mu_{X} $ which is defined through $ \lambda(\mu_{X})\define e^{4}(\mu_{X})\left(6-36 \log\left[e(\mu_{X})\right]\right) $, then we recover the result of\room\cite{Andreassen:2014eha},
\begin{align*}
V(\pbmin)=-&\hbar\frac{3}{8}e^{4}\mu_{X}^{4}\\*
+&\hbar^{2}e^{6}\mu_{X}^{4}\left(\frac{71}{6}-\frac{62}{3}\log{\left[e\right]}+10\log^{2}\!\left[e\right] \right).
\end{align*}
We also note that an alternative definition of $ \mu_{X} $ is that $ \pbh_{0}(\mu_{X})=\mu_{X} $. This is consistent with the procedure in\room\cite{Andreassen:2014eha}, where $ \mu_{X} $ is the scale around which the $ \hbar $-expansion is performed.

%% file: tex/discussion.tex
\section{Discussion}\label{sec:disc}
Perturbative calculations of $ \Vmin = V(\pbmin) $ face difficulties regarding gauge dependence and \ir{} divergences. In this paper we have argued that the $ \hbar $-expansion\room\cite{Patel:2011th} is capable of dealing with both of these issues. The key insight is that the separation of scales introduced in\room\cite{Elias-Miro:2014pca,Espinosa:2016uaw,Espinosa:2017aew} allows us to isolate the physical part of $ \Vmin $ as coming from hard momentum modes. This separation defines a new scale $ \pbh $, which is the minimum of the hard potential $ \VHard $. We note that $ \pbh $ is physical in the sense that it is gauge independent.

In previous attempts to understand the relation between gauge dependence and \ir{} divergences through the effective potential, special interest has been shown in the \cw{} model\room\cite{Andreassen:2014eha,Espinosa:2016uaw}. However, we find that this model is unique in the sense that it necessitates a modification of the perturbation expansion, by having the coupling $ \lambda $ scale as $ \lambda \sim \hbar $. The result is that a resummation seems to be necessary in order to properly include all the contributions at $ \hbar^{2} $, as was shown in\room\cite{Andreassen:2014eha}. Later it was then commented in\room\cite{Espinosa:2016uaw} that this resummation is precisely the same that removes the \ir{} divergences of the effective potential. We would like to comment on these two results, from the perspective of this paper.
 We expect that if one would calculate the full three loop effective potential in the \cw{} model, one would find that there are divergent contributions which the resummation procedure is not capable of dealing with. To remove them one would have to perform a consistent $ \hbar $ power counting as in\room\cite{Andreassen:2014eha}. However, we suggest that it is possible to calculate the $ \hbar^{3} $ term without resorting to resummation. The procedure would be to use the mass $ G=0 $ in the scalar propagators, and then also include $ \lambda $ insertions up to the proper order. This is equivalent to picking out the hard contribution to $ \Vmin $, and ignoring the soft contributions.

We find that the $ \hbar$-expansion yields a finite result up to as many orders as we are able to check\textemdash we show explicitly that they cancel up to three loops, and that the leading singularities cancel to all loops. We expect that the $\hbar$-expansion is finite to all orders.

That said, it is not clear \emph{why} the $ \hbar $-expansion would isolate the finite parts of $ \Vmin $. It is clear that the $ \hbar $-expansion gives a gauge invariant result of $ \Vmin $, hence gauge dependent singularities cancel. A tentative explanation would be that such \ir{} divergences are absent in the $ R_{\xi} $-gauges for finite $ \xi $, and are also spurious. However, note that our proof of the finiteness of $ \Vmin $ to three loops does not utilize the fact that we are dealing with a gauge theory. In fact, the proof would carry over to a global $ \text{O}\left(2\right) $ model (by taking the $ e \rightarrow 0 $ limit), in which the Goldstone bosons are not "eaten." Perhaps this theory is protected because it is possible to gauge it. This would pose interesting questions for the role of the Goldstone boson in models with spontaneously broken global symmetries.

There are a number of directions in which future research could go. One could try to systematically answer the question posed above, regarding why the $ \hbar $-expansion is finite. One possible venue of attack is to construct an explicit hard (or soft) EFT for Fermi gauge Abelian Higgs, in which the finiteness of $ \Vmin $ might be evident. Another direction is to try and extend the combinatorial machinery we developed, in order to show (or disprove) that $ \Vmin $ is finite to all orders. A complementary approach would be to calculate the 3-loop effective potential in the \cw{} model, in order to test our expectation that there are non-resummable \ir{} divergences, and that the correct physical contribution to $ \Vmin $ is only due to hard modes. Finally, it would be interesting to try and apply the hard/soft split in order to calculate other observables from the effective potential, such as the critical temperatures of phase transitions. By isolating the physical parts of $ V $ as coming from hard contributions, such calculations might be made simpler.

Recently, Martin and Patel calculated the 2-loop effective potential for a generalized gauge fixing, in which $ R_{\xi} $-gauges as well as Fermi gauges can be obtained by taking limits\room\cite{Martin:2018emo}. The authors calculate $ \Vmin $ by minimizing the effective potential numerically in Abelian Higgs as well as in the standard model. They find a $ \xi $-dependent result that diverges in the limit $ \xi \rightarrow \infty $, and suggest that this might be remedied through a resummation. It may well be that such a resummation exists, but as we have shown in this paper, the established resummation procedure for dealing with Goldstone \ir{} divergences is not consistent in these models. Nevertheless, it is possible to isolate the hard contributions to $ \Vmin $, such that the $ \xi \rightarrow \infty $ limit is safe (though we advocate that one should really use the $ \hbar $-expansion to calculate $ \Vmin $). With these caveats in mind, it could be interesting to explore the question whether it is possible to extend or modify the resummation procedure to also work for Fermi gauges.

%% file: tex/acknowledgments.tex
\vspace*{\leadingHeight}

\noindent \emph{Acknowledgments:} We thank Rikard Enberg and Gunnar Ingelman for helpful comments on the manuscript. The Feynman diagrams where drawn using \texttt{TikZ-Feynman}\room\cite{Ellis:2016jkw}.
\vspace{-\leadingHeight}

%% file: tex/conventions.tex
\section{Conventions and Feynman rules}\label{app:conventions}
\subsection{General conventions}
\vspace{-\leadingHeight}
\begin{itemize}
	\item We have rescaled $ \hbar $ with $ 16 \pi^{2} $ in order to simplify the formulas while still retaining $ \hbar $ as the loop-counting parameter.
	\item The metric is mostly minus: $ (+,-,-,-) $.
	\item The shorthand for momentum integration is
	\begin{equation*}
		\momint{q}=Q^{2\epsilon}\int \frac{\dif^{d}q}{\left(2 \pi\right)^{d}},
	\end{equation*}
	with $ Q $ the renormalization scale. We use $ \mu $ to denote the \msbar{} scale.
	\item The field-dependent masses squared are denoted by capital letters signifying the particle it corresponds to (e.g. the Higgs mass squared would be denoted $ H $).
	\item We use the shorthand $ \smash{\LOG{X}\define\log\left(X/\mu^{2}\right)}$.
	\item Derivatives with respect to the background field $ \pb $ are denoted $ \partial $. Spacetime derivatives are denoted as usual, with a Lorentz index $ \partial_{\mu} $. Any other derivatives are denoted with a subscript, e.g. $ \partial_{\xi}\define\partial/\partial\xi   $, except for renormalization scale derivatives which we explicitly spell out as $ \frac{\partial}{\partial \mu} $.
	\item $ \sun = \momint{k}(\dif l) (\dif q)\delta^{d}\left(k^{\mu}+q^{\mu}+l^{\mu}\right) $ denotes the momentum integration of the setting sun diagrams.
	\item We use $ \simeq $ to denote equivalence up to subleading singularities.
\end{itemize}
\subsection{Abelian Higgs Feynman rules}\label{sapp:feynrules}
Feynman rules in the presence of spontaneous symmetry breaking involve insertions of the background field $\pb$; it is possible to have arbitrary many $\pb$ insertions in a diagram. We follow the approach in\room\cite{Andreassen:2014eha}, where the effective potential is computed through Feynman diagrams with dressed propagators, propagators with all $\pb$ insertions included. The dressed propagators are found by diagonalizing the kinetic terms in the presence of the background field $ \pb $. In Abelian Higgs the dressed propagators are
\begingroup
\allowdisplaybreaks
\begin{alignat*}{3}
	&\begin{gathered}
		\includegraphics[width=0.1\textwidth]{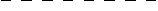}
	\end{gathered}
	&&=D_{G}(k)&&= i \frac{k^{2}-\xi A}{\left(k^{2}-G_{+}\right)\left(k^{2}-G_{-}\right)},\\
	&\begin{gathered}
		\includegraphics[width=0.1\textwidth]{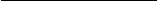}
	\end{gathered}
	&&=D_{H}(k)&&= i \frac{1}{k^{2}-H},\\
	&\begin{gathered}
		\includegraphics[width=0.1\textwidth]{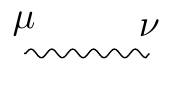}
	\end{gathered}
	&&=D_{\mu\nu}(k)&&= - i \left( \frac{1}{k^{2}-A}P^{\mu\nu}\right.\\*
	& && &&+\left.\xi \frac{k^{2}-G}{\left(k^{2}-G_{+}\right)\left(k^{2}-G_{-}\right)}\frac{k^{\mu}k^{\nu}}{k^{2}}\right),\\
	&\begin{gathered}
		\includegraphics[width=0.1\textwidth]{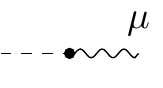}
	\end{gathered}
	&&=F^{\mu}(k)&&= \xi \sqrt{A} \frac{-k^{\mu}}{\left(k^{2}-G_{+}\right)\left(k^{2}-G_{-}\right)},\\
	&\begin{gathered}
		\includegraphics[width=0.1\textwidth]{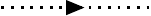}
	\end{gathered}
	&&=D_{c\overline{c}}(k)&&= i \frac{1}{k^{2}},
\end{alignat*}
\endgroup
where the momentum flows from left to right. The final propagator is the ghost propagator. The vertices in the presence of $ \pb $ are
\begingroup
\allowdisplaybreaks
\begin{alignat*}{3}
	\begin{gathered}
		\includegraphics[height=35pt]{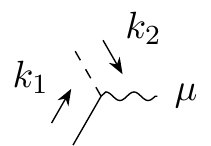}
	\end{gathered}&=e\left(k_{1}-k_{2}\right)^{\mu},\qquad
	&&\begin{gathered}
		\includegraphics[height=35pt]{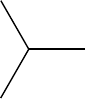}
	\end{gathered}&&=-i\lambda \pb,\\
	\begin{gathered}
		\includegraphics[height=35pt]{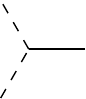}
	\end{gathered}&=-i\frac{\lambda}{3} \pb,
	&&\begin{gathered}
		\includegraphics[height=35pt]{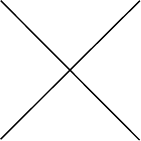}
	\end{gathered}&&=-i\lambda,\\
	\begin{gathered}
		\includegraphics[height=35pt]{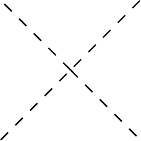}
	\end{gathered}&=-i\lambda,
	&&\begin{gathered}
		\includegraphics[height=35pt]{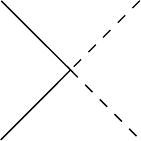}
	\end{gathered}&&=-i\frac{\lambda}{3},\\
	\begin{gathered}
		\includegraphics[height=35pt]{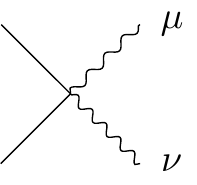}
	\end{gathered}&=2 i e^{2}g_{\mu\nu},
	&&\begin{gathered}
		\includegraphics[height=35pt]{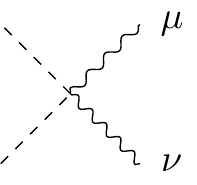}
	\end{gathered}&&=2 i e^{2}g_{\mu\nu},\\
	\begin{gathered}
		\includegraphics[height=35pt]{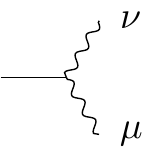}
	\end{gathered}&=2 i e^{2} \pb g_{\mu\nu}.&&
\end{alignat*}
\endgroup

%% file: tex/ward.tex
\section{Ward identity of the mixed propagator}\label{app:ward}
In our analysis of the Nielsen coefficient we utilized generalized Ward identities in order to parametrize the mixed $ (G,A^{\mu}) $ propagator in a useful way. The propagator Ward identity for Abelian higgs is well known and the derivation is given in appendix A.2 of\room\cite{Espinosa:2016uaw}. We find the approach in\room\cite{Espinosa:2016uaw} very convenient, and we follow along their derivation, except that we retain some extra terms that vanish when one takes the limit $ \pb \rightarrow \pbmin $.

With a slight abuse of notation the Ward identity for the matrix $ (G,A^{\mu}) $ matrix propagator $ \mathcal{D}(k) $ is given by
\begin{align}
\mathcal{D}(k)^{-1}&\begin{pmatrix}
i k_\mu \\ \sqrt{A}
\end{pmatrix}
+\begin{pmatrix}
\frac{k^2 k_\nu}{\xi}\label{eq:ward}\\
-i e \partial_{\phi} V
\end{pmatrix}
=0,\\
\mathcal{D}(k)^{-1}&\define\begin{pmatrix}
\left(\mathcal{D}_A^{\mu \nu}\right)^{-1}(k) & \left(\mathcal{D}^{\nu}_{A G}\right)^{-1}(k)\\
\left(\mathcal{D}^{\mu}_{G A}\right)^{-1}(k) &  \mathcal{D}_{G}^{-1}(k)
\end{pmatrix}.\nonumber
\end{align}
It is useful to decompose the propagator into a transverse and a longitudional part
\begin{align*}
\mathcal{D}(k)&=\begin{pmatrix}
P^{\mu \nu} & 0\\
0&0
\end{pmatrix}
G_T(k)\\*
&+
\begin{pmatrix}
i \frac{k_\nu}{\sqrt{k^{2}}} & 0\\
0&1
\end{pmatrix}
G_L(k)
\begin{pmatrix}
-i \frac{k_\mu}{\sqrt{k^{2}}} & 0\\
0&1
\end{pmatrix},
\end{align*}
where $ G_{T}(k) $ is a scalar function that gives the self-energy of the transverse modes, and $ G_{L}(k) $ is a $ 2 \times 2 $ matrix that does the same for the mixed longitudinal modes. Inserting this decomposition into the Ward identity\room\eqref{eq:ward}, one finds

\begingroup
\small
\begin{align*}
G_L^{-1}(k)=i \begin{pmatrix}
k^2/\xi &0 \\
0& \partial_{\phi} V
\end{pmatrix}
-i \Pi_{\textcaps{L}}(k)\begin{pmatrix}
A  & - \sqrt{A k^{2}}\\
- \sqrt{A k^{2}} & k^2
\end{pmatrix},
\end{align*}
\endgroup
where $ \Pi_{\textcaps{L}}(k)=1+\ordo(\hbar) $ is a scalar function that parametrizes the self energy of the longitudinal mixed modes. The propagator can now be written

\begingroup
\small
\begin{align*}
\mathcal{D}(k)=&\begin{pmatrix}
P^{\mu \nu} & 0\\
0&0
\end{pmatrix}
G_T(k)\\*
+i&\begin{pmatrix}
-\xi \left(k^2-\overline{G}/\Pi_{\textcaps{L}}(k)\right)\frac{1}{\widetilde{\Delta}(k)}\frac{ k^\mu k^\nu}{k^2} & -i \xi \sqrt{A} k^\nu/\widetilde{\Delta}(k) \\ i \xi \sqrt{A} k^\mu/\widetilde{\Delta}(k) & \frac{k^2-\Pi_{\textcaps{L}}(k) A \xi}{\Pi_{\textcaps{L}}(k) \widetilde{\Delta}(k)}
\end{pmatrix},
\end{align*}
\endgroup
where we introduced some shorthands. They are:
\begin{align*}
\overline{G}&=\frac{1}{\pb}\partial V,\\
\widetilde{G}_{\pm}&=\frac{\overline{G}_{\pm}}{\Pi_{\textcaps{L}}(k)},\\
\widetilde{\Delta}(k)&=\left(k^{2}-\widetilde{G}_{+}\right)\left(k^{2}-\widetilde{G}_{+}\right).
\end{align*}
In particular, this tells us that the exact mixed $ \left(G, A^{\mu}\right) $ propagator $ \overline{F}^\mu(k) $ can be written as
\begin{equation*}
	\overline{F}^\mu(k)= \xi \sqrt{A}\frac{-k^\mu}{\left(k^{2}-\widetilde{G}_{+}\right)\left(k^{2}-\widetilde{G}_{-}\right)},
\end{equation*}
which is the form we utilized in subsubsection\room\ref{sssec:nielsen}.

%% file: tex/2LoopPot.tex
\section{2-loop effective potential}\label{app:2loop}
In this appendix we collect all the Feynman diagrams required for calculating the 2-loop effective potential of Fermi gauge Abelian Higgs. We do not write their explicit values because these are often not very illuminating. We do however write them in terms of symmetry factors, coupling constants, and integrals over propagators. There are two classes of integrals, the double bubbles and the setting suns. Because we use the same metric as in\room\cite{Andreassen:2014eha}, we can use their master integrals for computing these diagrams. Note that certain diagrams require reduction before the master integrals are applicable. This can be performed by hand, using tricks such as partial fraction decomposition, or it can be automated by using tools such as \texttt{FIRE}\room\cite{Smirnov:2014hma}.
\subsection{Double bubbles}
The double bubbles can be thought of as a 1-loop vacuum integral squared, times an appropriate coupling constant and symmetry factor. The diagrams in this class are
\begin{center}
	\begin{tabular}{ll}
		$ \	\mathcal{M}_{\text{A}} = 	\begin{gathered}
		\includegraphics[width=0.1\textwidth]{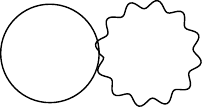}
		\end{gathered}\room, $ &
		$ \mathcal{M}_{\text{B}} = 	\begin{gathered}
		\includegraphics[width=0.1\textwidth]{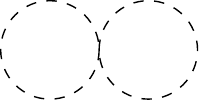}
		\end{gathered}\room, $\\
		\vspace{-1em}& \\
		$ \mathcal{M}_{\text{C}} = 	\begin{gathered}
		\includegraphics[width=0.1\textwidth]{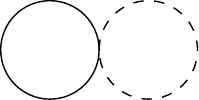}
		\end{gathered}\room, $
		&
		$ \mathcal{M}_{\text{D}} = 	\begin{gathered}
		\includegraphics[width=0.1\textwidth]{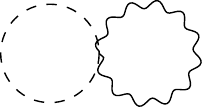}
		\end{gathered}\room, $\\
		\vspace{-1em}& \\
		$ \mathcal{M}_{\text{E}} = 	\begin{gathered}
		\includegraphics[width=0.1\textwidth]{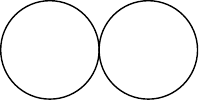}
		\end{gathered}\room, $
		&
		$ \mathcal{M}_{\text{F}} = \begin{gathered}
		\includegraphics[width=0.1\textwidth]{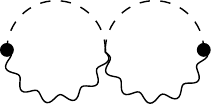}
		\end{gathered}\room. $
	\end{tabular}
\end{center}
In terms of symmetry factors, coupling constants, and momentum integrals, these diagrams are given by
\begingroup
\allowdisplaybreaks
\begin{alignat*}{2}
	\mathcal{M}_{\text{A}} &= 	\begin{gathered}
		\includegraphics[height=30pt]{EffPotHbExpPDF-figure42}
	\end{gathered} &&= 2 i^2 e^{2}\frac{1}{2^2}\times\\*
	& &&\times\momint{k}D_{H}(k)\momint{q} \mathcal{D}^{\mu}_{\mu}(q),\\
	\mathcal{M}_{\text{B}} &= 	\begin{gathered}
		\includegraphics[height=30pt]{EffPotHbExpPDF-figure43}
	\end{gathered} &&= -i^2\lambda\frac{1}{2^3}\left(\momint{q}D_{G}(q)\right)^{2},\\
	\mathcal{M}_{\text{C}} &= 	\begin{gathered}
		\includegraphics[height=30pt]{EffPotHbExpPDF-figure44}
	\end{gathered} &&= i\frac{-i\lambda}{3}\frac{1}{2^2}\times\\*
	& &&\times\momint{k}D_{H}(k)\momint{q}D_{G}(q),\\
	\mathcal{M}_{\text{D}} &= 	\begin{gathered}
		\includegraphics[height=30pt]{EffPotHbExpPDF-figure45}
	\end{gathered} &&= 2 i^2 e^{2}\frac{1}{2^2}\times\\*
	& &&\times\momint{k}D_G(k)\momint{q} \mathcal{D}^{\mu}_{\mu}(q),\\
	\mathcal{M}_{\text{E}} &= 	\begin{gathered}
		\includegraphics[height=30pt]{EffPotHbExpPDF-figure46}
	\end{gathered} &&= i\frac{-i\lambda}{2^3}\left(\momint {k}D_H(k)\right)^2,\\
	\mathcal{M}_{\text{F}} &= \begin{gathered}
		\includegraphics[height=30pt]{EffPotHbExpPDF-figure47}
	\end{gathered} &&= 0.
\end{alignat*}
\endgroup
The diagram $ \mathcal{M}_{\text{F}} $ vanishes by symmetric integration.
\subsection{Setting suns}\label{Sec:Sunset}
The setting suns are slightly more complicated because the two momentum integrals cannot be separated. We use the shorthand $ \sun $ to represent the momentum integration $ \smash{\momint{k}(\dif l) (\dif q)\delta^{d}\left(k^{\mu}+q^{\mu}+l^{\mu}\right)} $. In the following table of setting sun diagrams, we have included a diagram with two ghost propagators, even though in Fermi gauges this diagram is not present (ghosts are free).
\begin{center}
	\begin{tabular}{lll}
		$ \	\mathcal{M}_{\text{G}} = \begin{gathered}
		\includegraphics[height=30pt]{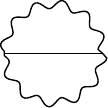}
		\end{gathered}\room, $ &
		$ \	\mathcal{M}_{\text{H}} = \begin{gathered}
		\includegraphics[height=30pt]{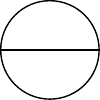}
		\end{gathered}\room, $ &
		$ \	\mathcal{M}_{\text{I}} = \begin{gathered}
		\includegraphics[height=30pt]{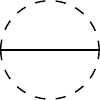}
		\end{gathered}\room, $\\
		\vspace{-1em}& & \\
		$ \mathcal{M}_{\text{J}} = \begin{gathered}
		\includegraphics[height=30pt]{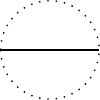}
		\end{gathered} \room, $ &
		$	\mathcal{M}_{\text{K}} = \begin{gathered}
		\includegraphics[height=30pt]{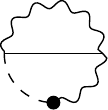}
		\end{gathered}\room, $ &
		$ \mathcal{M}_{\text{L}} = \begin{gathered}
		\includegraphics[height=30pt]{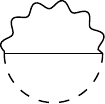}
		\end{gathered}\room, $ \\
		\vspace{-1em}& & \\
		$ \mathcal{M}_{\text{M}} = \begin{gathered}
		\includegraphics[height=30pt]{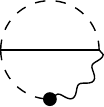}
		\end{gathered}\room, $ &
		$ \mathcal{M}_{\text{N}} = \begin{gathered}
		\includegraphics[height=30pt]{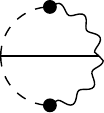}
		\end{gathered}\room, $ & $ \mathcal{M}_{\text{O}} =
		\begin{gathered}
		\includegraphics[height=30pt]{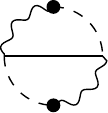}
		\end{gathered}\room. $
	\end{tabular}
\end{center}
In terms of symmetry factors, coupling constants, and momentum integrals, the individual diagrams are

\begingroup
\small
\allowdisplaybreaks
\begin{alignat*}{2}
	\mathcal{M}_{\text{G}}&=
	\begin{gathered}
		\includegraphics[height=30pt]{EffPotHbExpPDF-figure60}
	\end{gathered}&&= i\frac{1}{4} (2ie^2\phi)^2\times\\*
	& &&\times \sun D_H(k) \mathcal{D}^{\mu \nu}(q) \mathcal{D}_{\mu \nu}(l),\\
	\mathcal{M}_{\text{H}}&=
	\begin{gathered}
		\includegraphics[height=30pt]{EffPotHbExpPDF-figure61}
	\end{gathered}&&=i\frac{1}{2 \times 3!}(-i\lambda \phi)^2\times\\*
	& && \times \sun D_H(k)D_H(k) D_H(k),\\
	\mathcal{M}_{\text{I}}&=\begin{gathered}
		\includegraphics[height=30pt]{EffPotHbExpPDF-figure62}
	\end{gathered} && = \frac{i}{4}\left(\frac{-i\lambda \phi}{3}\right)^2 i^3\times\\*
	& && \times \sun D_H(k) D_G(l)D_G(q),\\
	\mathcal{M}_{\text{J}}&=
	\begin{gathered}
		\includegraphics[height=30pt]{EffPotHbExpPDF-figure63}
	\end{gathered} && = 0,\\
	\mathcal{M}_{\text{K}}&=
	\begin{gathered}
		\includegraphics[height=30pt]{EffPotHbExpPDF-figure64}
	\end{gathered} && = i e(2 i e^2 \phi)\times\\*
	& && \times \sun D_H(k)(k-l)^\mu \mathcal{D}_{\mu \nu}(q)F^\nu(l),\\
	\mathcal{M}_{\text{L}}&=
	\begin{gathered}
		\includegraphics[height=30pt]{EffPotHbExpPDF-figure65}
	\end{gathered} && = \frac{i}{2}e^2(-1)\times\\*
	& && \!\times \sun (k-l)^\mu \mathcal{D}_{\mu \nu}(q) (k-l)^\nu D_H(k) D_G(l), \\
	\mathcal{M}_{\text{M}}&=
	\begin{gathered}
		\includegraphics[height=30pt]{EffPotHbExpPDF-figure66}
	\end{gathered} && = i(-i\frac{\lambda \phi}{3})e \times\\*
	& && \times \sun D_H(k) D_G(l)(l-k)^\mu F_\mu(q),\\
	\mathcal{M}_{\text{N}}&=
	\begin{gathered}
		\includegraphics[height=30pt]{EffPotHbExpPDF-figure67}
	\end{gathered} && = \frac{i}{2}(2 i e^2 \phi)(-i \frac{\lambda \phi}{3}) \times\\*
	& && \times\sun D_H(k)F^\mu(l)F_\mu(q),\\
	\mathcal{M}_{\text{O}}&=
	\begin{gathered}
		\includegraphics[height=30pt]{EffPotHbExpPDF-figure68}
	\end{gathered} && = \frac{i}{2}e^2 \times\\*
	& && \times \sun D_H(k)((l-k)^\mu F_{\mu}(q) (q-k)^\nu F_{\nu}(l).\\
\end{alignat*}
\endgroup

%% file: tex/LDS.tex
\section{Cancellation of certain subleading singularities}\label{sec:LDS}
In subsubsection\room\ref{sssec:irfit} we have showed that the leading singularities of $ \Vmins $ cancel to all loop orders. Because the 3-loop potential only has leading singularities, this means that the 3-loop potential is finite. We demonstrated this fact explicitly. We are not able to show that all singularities cancel, but in this appendix we show that certain classes of subdivergences cancel to all orders. This involves non-trivial combinatorics, and we see this as further motivation that $ \Vmin $ is finite in the $ \hbar $-expansion.

We will only consider singularities coming from diagrams of the type
\begin{equation*}
	\begin{gathered}
		\includegraphics[width=0.1\textwidth]{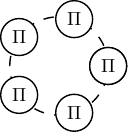}
	\end{gathered}\room,
\end{equation*}
where the self-energy $\Pi(k^2)\sim\hbar ~\Pi_1(k^2)+\hbar^2~\Pi_2(k^2)+\ldots$ only contains hard contributions. The momentum dependence of $ \Pi(k^{2}) $ will soften the divergence, because for soft momenta $ \Pi(k^{2})=\Pi(0)+G \Pi^{(1)}(0)+\mathellipsis $ The most divergent terms correspond to just using $ \Pi(0) $ in the self-energy insertions. We showed that leading singularities associated with such insertions cancel in subsubsection\room\ref{sssec:irfit}. That the singularities with the momentum dependent self-energy $ \Pi(k^{2}) $ cancel in the $ \hbar $-expansion is not clear \emph{a priori}, because $ \Pi^{(1)}(0) $ and higher derivatives correspond to contributions with non-zero external momentum --- which cannot be obtained from the effective potential.

In subsection\room\ref{ss:lsmomdep} we show that given a specific insertion in a Daisy diagram, the leading divergences with respect to this insertion cancel. In subsection\room\ref{ss:subleading} we show that a certain class of subleading divergences cancel.

\subsection{Leading singularities and momentum dependent insertions}\label{ss:lsmomdep}
In this subsection we will prove that given a choice of insertions in a daisy diagram, the leading singularities cancel. We will do this in multiple steps: cancellation of singularities with $\Pi_1(0)$ and $\Pi^{(1)}_1(0)$ insertions, cancellation of singularities with $\Pi_1(k^2)$ insertions, and finally cancellation of singularities with $\Pi(k^2)$ insertions.
\subsubsection{ $\Pi_{1}^{(1)}$ insertions}
We now consider leading daisy diagrams of loop level $ l $ of which $i$ of the petals are $\smash{k^2 \Pi_{1}^{(1)}(0)}$ insertions. These diagrams are less and less divergent with increasing $ i $, which means that we can consider them separate from each other. In this subappendix we will consider such an insertion as given, and then count the leading singularities for this insertion. Denoting such a diagram by $ V_l^{i} $, we have that
\begin{align*}
V_l^{i}\simeq \binom{l-1}{i}\frac{1}{l-1}(-1)^{l-1}& (\Pi_1(0))^{l-1-i}(\Pi_{1}^{(1)}(0))^i\times\\*
\times & \momint{k}\frac{(k^2)^i}{(k^2+G)^{l-1}}.
\end{align*}
We now use the relation
\begin{equation*}
\frac{x^i}{(x+y)^n}=\sum_{k=0}^{i}\binom{i}{k}(-1)^{k}\frac{y^k}{(x+y)^{n+k-i}}
\end{equation*}
and equation \ref{eq:ldaisyDeriv} to rewrite $V_l^{i}$ as
\begingroup
\begin{align*}
V_l^{i}\simeq  \binom{l-1}{i}\frac{(-1)^{i}}{l-1}&(\Pi_1(0))^{l-1-i}(\Pi_{1}^{(1)}(0))^i\times \\*
&\times \sum_{k=0}^{i}\binom{i}{k}G^k\frac{\partial_{G}^{l+k-i-1}\VSoft_1}{(l+k-i-2)!}.
\end{align*}
\endgroup
In the $\hbar$-expansion a generic $ V_{l}^{i} $ diagram will contribute through $\partial^n V_l^{i}$. With this in mind the goal is to convert the $ G $-derivatives to $ \pb $-derivatives; we note that at the singularity order we are considering, $\partial^n=\left(\partial^2 V_0/\phi_0\right)^n \partial_{G}^n$. We also need the identity
\begin{align}\label{eq:Leibniz}
\partial_x^n (x^k V(x))=\sum_{j=0}^k \binom{k}{j}\binom{n}{j}j!~ x^{k-j}\partial_x^{n-j}V(x),
\end{align}
where $ V(x) $ is a generic function. Using equation \ref{eq:Leibniz}, and suppressing factors independent of $ n $ and $ l $, we can then write $\partial^n V_l^{i}$ as
\begin{align*}
\partial^n V_l^{i}&\simeq \binom{l-1}{i}\frac{1}{l-1}\times\\*
&\times\sum_{k=0}^i \sum_{j=0}^k \binom{i}{k}\binom{k}{j}\binom{n}{j}j!G^{k-j} \frac{\partial^{l+k+n-i-j-1}\VSoft_1}{(l+k-i-2)!}.
\end{align*}
Each instance of $ \partial^n V_l^{i} $ will contribute in the sum over $ n $, as in equation\room\eqref{eq:VminSL}. With this in mind, let us try to find a more tractable expression for the double sum. By changing the summation variables, the sum takes the form
\begin{align*}
\sum_{m=0}^i \sum_{j=0}^{i-m} \binom{i}{j+m}\binom{j+m}{j}\binom{n}{j}j!G^{m} \frac{\partial^{l+m+n-i-1}V_1}{(l+j+m-i-2)!}.
\end{align*}
Fixing $m$ we can perform the inner sum,
\begin{align*}
&\sum_{j=0}^{i-m} \binom{i}{j+m}\binom{j+m}{j}\binom{n}{j}j! \frac{1}{(l+j+m-i-2)!}=\\*
&\frac{1}{(l-2)!}\binom{i}{i-m}\frac{(l+n-2)!}{(l+n+m-i-2)!}.
\end{align*}
We now have all ingredients we need to perform the sum in equation\room\eqref{eq:VminSL}. Following the same steps as in subsubsection\room\ref{sssec:irfit}, we find that the contribution of leading singularities to the $ \hbar^{L} $ term from daisy diagrams with $ i $ insertions of $ k^{2} \Pi_{1}^{(1)}(0) $ is
\begin{align}\label{eq:LeadingSingCancel}
\begin{split}
&\sum_{n=0}^{L-i-1}\frac{\phi^n \partial^n V^{i}_{L-n}}{n!}\simeq\\*
&\sum_{m=0}^{i}(\text{overall factor}) \sum_{n=0}^{L-i-1}\frac{(-1)^n}{n! (L-1-i-n)!}.
\end{split}
\end{align}
This sum vanishes for all $ m $ and $ i $ because of equation \ref{eq:Binomial}.

\subsubsection{Cancellation of full $\Pi_{1}(k^2)$ insertions}
In light of the previous subsubsection, we are ready to show the cancellation of leading singularities given full $\Pi_1(k^2)$ insertions.

Consider a daisy diagram with $a_1$ insertions of $\Pi_1^{(1)}(0) k^2$, $a_2$ insertions of $\smash{\Pi_1^{(2)}(0) (k^2)^2}$, ..., $a_\alpha$ insertions of $\smash{\Pi_1^{(\alpha)}(0) (k^2)^\alpha}$. All such diagrams have a common factor from the Taylor expansion of $\Pi_1(k^2)$, the only difference is the combinatorial factor from the insertions. A general diagram is given by
\begin{align*}
V_l^{\bar{i}}&\simeq  \frac{(l-2)!}{a_1!\ldots a_\alpha! (l-1-\bar{i})!} (\Pi_1(0))^{l-1-\bar{i}}\times\\*
& \times \prod^\alpha_{q} (\Pi_1(0)^{(q)})^{a_q}\momint{k}\frac{(k^2)^{\bar{i}}}{(k^2+G)^{l-1}},
\end{align*}
where $\bar{i}\equiv \sum_{q}^\alpha a_q$. Up to a factor common for all diagrams the result is the same as in the previous section with $i\rightarrow \bar{i}$, and the singularities vanish by the same arguments.

We have now shown that all leading singularities with $\Pi_1(k^2)$ insertions vanish to all orders in $\hbar$. Note that the explicit form of $\Pi_{1}(k^2)$ was not important for the proof; the only relevant fact was that $\Pi_{1}(k^2)$ is finite and can be expanded in powers of $k^2$.
\subsubsection{Cancellation of $\Pi(k^2)$ insertions}
For the final step we have to deal with insertions of higher order self-energies such as $\Pi_{2}(0)$. There is a slight complication with these insertions because there are different families of terms in the $\hbar$-expansion that are of the same singularity order. At four loops, including only terms that contain $\Pi_2(0)$, this is reflected in the $ \hbar $-expansion as
\begin{align*}
V(\pbmin)=\mathellipsis+\hbar^{4}&\left(V_4+\phi_{1} \partial V_3+\phi_{1}^2 \partial^2 V_2/2 \right.\\*
&\left.+\phi_{2} \partial V_2+\phi_1 \phi_{2} \partial V_2 \right)+\mathellipsis
\end{align*}
The first family corresponds to terms with the maximal number of $ \pb_{1} $ factors, here $ \phi_{1}^2 \partial^2 V_2/2  $. However, note that for the $\phi_{2} \partial V_2$ term to be of the same singularity order as $V_4$ we need $\partial V_2$ to be leading order divergent. We can generalize this statement to terms with arbitrary many factors of $ \pb_{q},\room q>1 $.

First, note that at loop order $ L $ all terms with a factor $\pb_{2}^i$ will factorize as
\begin{align*}
\sum_{n=0}^{L-2i-1}\frac{\phi_2^i}{i!}\left(\frac{\phi_1^n \partial^{i+n}V_{L-2i-n}}{n!}\right).
\end{align*}
This sum is of the same type as in equation \ref{eq:LeadingSingCancel}, with $L\rightarrow L-2i$, and the sum thus vanishes by the same arguments. The second and final step in the generalization is that the same statement is true for arbitrary insertions of $\phi_q$. The result factorizes as
\begin{align*}
\sum_{n=0}^{L-1-\bar{q}}\prod_{q=0}^{\alpha} \frac{\phi_q^{a_q}}{a_q!}\left(\frac{\phi_1^n \partial^{n+\bar{i}}V_{L-n-\bar{q}}}{n!}\right),
\end{align*}
where we have defined $\bar{q}=\sum_{i=2}^{\alpha} q a_q,~\bar{i}=\sum_{i=2}^{\alpha} a_q$. As long as we are considering leading singularities for a given insertion, the object in the parenthesis must be leading order. Again, the sum over $n$ vanishes by the same arguments as in equation \ref{eq:LeadingSingCancel}.

The only remaining leading daisy diagrams in the $\hbar$ expansion come from the first family of terms of the form $\sum_n \frac{1}{n!}\phi_1^n \partial^n V$.  Let us first focus on the leading singularities for diagrams $V_l^{i}$ with $i$ insertions of $\Pi_{2}(0)$,
\begin{align*}
V_l^{i}&\simeq \binom{l-1-i}{i}\frac{(-1)^{l-1-i}}{l-1-i}(\Pi_{2}(0))^i (\Pi_{1}(0))^{l-1-2i}\times\\*
&\times\momint{k}\frac{1}{(k^2+G)^{l-1-i}}.
\end{align*}
Using the by now familiar bag of tricks we find
\begin{align*}
\sum_{n=0}^{L-2i-1} \frac{\partial^n V_{L-n}}{n!}\propto  \sum_{n=0}^{L-2i-1}\frac{(-1)^n}{n!(L-2i-1-n)!}=0,
\end{align*}
by equation \ref{eq:Binomial}.

Insertions of an arbitrary number of $(\Pi_{\alpha}(0))$ will not change the combinatorics pertinent to the sum in the $ \hbar $-expansion, just as in the previous subsubsection. This again leads to the same sum, which vanishes. Note that by the same argument, taking into account the subleading $k^2$ dependence of the self-energies will end up with the same sum in the end (modulo overall factor), and we have thus proved that given insertions of $ \Pi(k^{2}) $, the leading singularities cancel to all orders in pertubation theory.

\subsection{Subleading singularities}\label{ss:subleading}
In this subsection we discuss a particular class of subleading singularities. Even though we cannot prove that all subleading divergences cancel, we argue that the non-trivial combinatorics needed for this cancellation is further support for the finiteness of $ \Vmin $.

We will focus on subleading singularities arising from daisy diagrams with $\Pi_{1}(0)$ insertions, and no derivatives acting on $\partial \VHard_{1}$. In the $ \hbar $-expansion this corresponds to the terms with $ \pb_{1} $ and $ \pb_{2} $ factors. We refer to these as two families: the terms with only $ \pb_{1} $ factors, and the terms with both $ \pb_{1} $ and $ \pb_{2} $ factors. To translate the daisy diagrams to these kinds of terms we follow the procedure laid out in the previous subsection, but now we need to retain subleading divergences.\footnote{In this subsection we hence let $ \simeq $ represent equivalence up to subsubleading divergences.} With this in mind, we rewrite the $ l $-loop level daisy $ \VSoft_{l} $ in terms of $ \pb $-derivatives acting on $ \VSoft_{1} $; we need the following two relations:
\begin{align*}
&(\partial_{G})^{l-1}=\left(\frac{3}{\lambda}\right)^{l-1}\sum_{k=0}^{l-2}\frac{(l+k-2)!}{2^k(l-k-2)!k!\phi^{l+k-1}}(-1)^{k}\partial^{l-k-1},
\\& \partial^n (\frac{1}{x^k}V(x))=\sum_{j=0}^{n}\frac{1}{x^{k+j}}(-1)^j\binom{k+j-1}{j}\binom{n}{j}j! \partial^{n-j} V(x).
\end{align*}
A general term in the $ \hbar $-expansion is of the form $ \partial^{n} \VSoft_{l}$. To find an expression for this quantity, note that because we focus on daisy diagrams with $\Pi_{1}(0)=\partial \VHard_{1}/\phi$ insertions, every $\Pi_{1}(0)$ insertion contributes a factor $\phi^{-1}$.

 This fact, together with the two formulas above, yields the expression
\begin{align*}
\partial^n \VSoft_l &\simeq (\phi_{1})^{l-1} \sum_{k=0}^{l-2}\sum_{j=0}^{n}(-1)^{l+k+j	-1}\frac{(l+k-2)!}{2^k(l-k-2)!k!(l-1)!}
\\&\times\binom{k+j+2l-3}{j}\binom{n}{j}j!\partial^{l+n-k-j-1}\VSoft_1.
\end{align*}
According to the standard procedure, we will sum over $ n $ in the $ \hbar $-expansion, and could use a more tractable expression for $ \partial^n V_l $. To achieve this we want to isolate specific values of $\partial^x V_1$; we rewrite the sum above as

\begingroup
\small
\begin{align}\label{eq:subleading}
\begin{split}
  \partial^n \VSoft_l &\simeq (\phi_{1})^{l-1} \sum_{m=0}^{L-1}\sum_{j=0}^{m}(-1)^{L+m+n}2^{1+j-m}\frac{(L+m-n-j-3)!}{(m-j-1)!(L-n-1)!}\\ &\times\binom{n}{j}j!\binom{2L+m-2n-4}{j}\frac{1}{(j+L-m-n-1)!}\partial^{L-m}\VSoft_1,
\end{split}
\end{align}
\endgroup%
where we have used $l=L-n$, anticipating the sum over $n$.

We cannot perform the sum in equation \ref{eq:subleading}, but it turns out that we do not have to. Let $ a_{m}^{n} $ denote the coefficient of $ \partial_{L-m}\VSoft_{1} $ in $ \partial^{n} \VSoft_{l} $, and $ P(n,q) $ denote a polynomial in $ n $ of degree $ q $. Then, fixing $m$ and performing the sum over $j$, we find
\begin{align}
&a_{1}^{n}\sim (-1)^n\frac{1}{(L-n-1)!}\label{eq:a1},
\\&a_{m\geq 2}^{n}\sim (-1)^n\frac{P(n,2m-3)}{(L-n-2)!}\label{eq:a2}.
\end{align}
Even though we cannot calculate the coefficients of the polynomial in $ a_{m\geq 2}^{n} $, it is still possible to show that the subleading divergences cancel. The $ \hbar^{L} $ term in the $ \hbar $-expansion now takes the form
\begin{equation*}
	\sum_{n=0}^{L-2}\pb_{1}^{n} \frac{\partial^n \VSoft_{L-n}}{n!}\simeq (\text{coeff.}) \sum_{n=0}^{L-2}\pb_{1}^{n}\sum_{m=1}^{L-1} a_{m}^{n}\frac{1}{n!} \partial^{L-m} \VSoft_{1},
\end{equation*}
 The $ m=1 $ terms are just the leading singularities $ \partial_{L-1} \VSoft_{1} $, which we have already shown to cancel.

All that remains are the $ m\geq 2 $ terms. We can eliminate some of them with the help of the relation\footnote{This and similar formulas can be proved by taking derivatives of $(1-x)^c$.}
\begin{align}\label{eq:masterBinomial}
\sum_{n=0}^{c}(-1)^n\binom{c}{n}n^q=
\begin{cases}
0       & \text{if } q<c \\
c!(-1)^c  & \text{if } q=c
\end{cases}.
\end{align}
Using the form of $ a_{m\geq 2}^{n} $ given in equation\room\eqref{eq:a2}, together with the relation\room\eqref{eq:masterBinomial}, we can eliminate all terms with $ m < \frac{L+1}{2} $.

What about the remaining terms? First, terms with $ m > \frac{L+1}{2} $ are beyond the scope of this paper. What remains is the edge case $ m = \frac{L+1}{2} $, which is only possible for odd loop levels. As it turns out, it is possible to determine the coefficient of the leading term in the polynomial $ P(n,q) $ present in equation\room\eqref{eq:a2}, that is
\begin{equation}\label{eq:slfamily1}
P(n,2m-3)\sim \left(\frac{3}{2}\right)^{m-1}\frac{1}{(m-1)!}n^{2m-3}
\end{equation}
We then have that
\begin{equation*}
\sum_{n=0}^{L-2}\pb_{1}^{n} \frac{\partial^n \VSoft_{L-n}}{n!}\simeq (\text{coeff.}) \pb_{1}^{L}\partial_{\frac{L-1}{2}} \VSoft_{1}\left(\frac{3}{2}\right)^{\frac{L-1}{2}} (-1)^{L-2}.
\end{equation*}
With the first family of terms taken care of, the remaining family consists of terms with $ \pb_{2} $ factors. The contribution of subleading divergences from both of these families to the $ \hbar^{L} $ term can be written in a compact form as
\begin{align*}
&\sum_{j=0}^{\frac{L-3}{2}}\frac{\pb^{j}_{2}}{j!}\sum_{n=0}^{L-1}\frac{\phi^n_{1}\partial^{n+j}V_{L-n-2j}}{n!}+\frac{\pb^{\frac{L-1}{2}}_{2}}{(\frac{L-2}{2})!}\partial^{\frac{L-2}{2}}\VSoft_{1}
\end{align*}
Performing the sum over $ n $ analogously to in equation\room\eqref{eq:slfamily1} and neglecting an overall pre-factor $\partial^{\frac{L-1}{2}}V_1$, we find
\begin{align*}
&\simeq \sum_{j=0}^{\frac{L-3}{2}} \frac{\phi^j_{2}\phi^{L-1-2j}_{1}}{j!}\frac{3^{\frac{L-1}{2}-j}}{(\frac{L-1}{2}-j)!2^{\frac{L-1}{2}-j}}+\frac{\pb^{\frac{L-1}{2}}_{2}}{(\frac{L-1}{2})!}=\\*
&\left(\frac{\left(3+2\frac{\pb_{2} \pb_0}{\pb^2_1}\right)}{2}\right)^{\frac{L-1}{2}}\frac{1}{(\frac{L-1}{2})!}.
\end{align*}
Finally, by solving the minimization condition we can translate $ \pb_{2} $ into powers of $ \pb_{1} $, in order to isolate the singularities we are interested in. We have that
\begin{align*}
\phi_{2}\partial^2 V_0+\partial V_2+\partial^2 V_1\phi_{1}+\partial^3 V_0 \frac{\phi^2_{1}}{2}=0 \implies \phi_{2}\sim -\frac{3}{2}\frac{\phi^2_{1}}{\phi_0},
\end{align*}
and thus the sum vanishes.

A corollary to the cancellation of this class of subleading divergences is that $\pbs$ is finite up to $\pbs_{3}$. This supports the observation made in\room\cite{Martin:2014bca}, which is that $ \pbmin $ is finite up to three loops in the standard model.